\documentclass[a4paper,11pt]{article}
\usepackage{jcappub} 
\usepackage{float}
\usepackage{cancel}
\usepackage{upgreek}
\usepackage{siunitx}
\usepackage{placeins}
\usepackage{orcidlink}
\usepackage{comment}
\usepackage[normalem]{ulem}
\usepackage{booktabs}
\usepackage{appendix}
\usepackage{indentfirst}
\usepackage{xcolor}
\usepackage{amsmath,mathtools}
\usepackage[T1]{fontenc} 
\usepackage{hyperref}
\usepackage{tikz}
\usepackage{booktabs}

\title{Cosmic voids evolution in modified gravity via hydrodynamics}

\author[a,b]{Tommaso Moretti,\orcidlink{0009-0006-4815-4764}}
\author[a,b]{Noemi Frusciante, \orcidlink{0000-0002-7375-1230}}
\author[c,d]{Giovanni Verza,\orcidlink{0000-0002-1886-8348}}
\author[e,f,g]{Francesco Pace \orcidlink{0000-0001-8039-0480}}

\affiliation[a]{Dipartimento di Fisica “E. Pancini”, Università degli Studi di Napoli “Federico II”,\\Compl. Univ. di Monte S. Angelo, Edificio G, Via Cinthia, I-80126, Napoli, Italy}

\affiliation[b]{INFN Sezione di Napoli, Università degli Studi di Napoli “Federico II”,\\Compl. Univ. di Monte S. Angelo, Edificio G, Via Cinthia, I-80126, Napoli, Italy}

\affiliation[c]{ICTP, International Centre for Theoretical Physics, Strada Costiera 11, 34151, Trieste, Italy}

\affiliation[d]{Center for Computational Astrophysics, Flatiron Institute, 162 5th Avenue, 10010, New York, NY, USA}

\affiliation[e]{Dipartimento di Fisica, Università degli Studi di Torino, Via P. Giuria 1, I-10125 Torino, Italy}

\affiliation[f]{INFN-Sezione di Torino, Via P. Giuria 1, I-10125 Torino, Italy}

\affiliation[g]{INAF-Istituto Nazionale di Astrofisica, Osservatorio Astrofisico di Torino,
strada Osservatorio 20, 10025, Pino Torinese, Italy}

\emailAdd{tommaso.moretti.pv@gmail.com}

\abstract{
  We present a hydrodynamical description of isolated spherical voids in modified gravity (MG),  extending the standard General Relativity (GR) and dynamical dark energy treatment by encoding gravity modifications into effective couplings that enter the Euler and Poisson equations. This yields a compact non-linear evolution equation for the Eulerian density contrast, controlled by a time- and density-dependent effective gravitational strength, and provides a direct map between model functions and void observables. We apply the framework to the luminal Galileon class of models, where derivative self-interactions generate Vainshtein screening and might lead to a breakdown of the physical branch in sufficiently underdense regions. Exploiting this feature, we apply  the void-informed viability requirement that translates into bounds on the theory parameter space and, equivalently, on the minimum attainable void depth as a function of redshift. For viable parameters of a concrete model, we quantify the impact of MG on isolated void evolution, the Lagrangian to Eulerian mapping, and the shell-crossing threshold. Relative to GR, we find a clear hierarchy of MG effects, with ${\cal O}(10\%)$ modifications in the gravitational couplings, percent-level shifts in the void density evolution, and sub-percent deviations in both the mapping and the shell-crossing thresholds. Moreover, within the adopted parametrization, we show analytically that voids always lie in an unscreened regime on the physical branch. Overall, the formalism provides a self-consistent route to predict void dynamics and consistency constraints in a broad class of MG models.
}

\begin{document}
\maketitle

\section{Introduction}

The $\Lambda$CDM model, resting on General Relativity (GR) plus a cosmological constant ($\Lambda$), has long served as the standard framework in cosmology, successfully accounting for the accelerated expansion of the Universe, the Cosmic Microwave Background (CMB) anisotropies, large-scale structure (LSS), and many other observations.  Nonetheless, in recent years, several persistent cosmic tensions have emerged, suggesting that $\Lambda$CDM may be incomplete. Chief among these are the Hubble tension~\cite{Breuval:2024lsv,Planck:2018vyg,SPT-3G:2022hvq,AtacamaCosmologyTelescope:2025blo,eBOSS:2020yzd,Zhang:2021yna,DESI:2025zgx,Freedman:2024eph,Scolnic:2023mrv,Planck:2016tof,Riess:2019cxk,Peterson:2021hel,Riess:2024ohe,Wojtak:2024mgg}, the disagreement between the locally measured value of the Hubble constant $H_0$ and its inference from early‐Universe data such as the CMB, and the $S_8$ tension~\cite{Asgari:2019fkq,KiDS:2020suj,Philcox:2021kcw,DES:2021bvc,Miyatake:2021sdd,Terasawa:2024agq,Planck:2018vyg,ACT:2020gnv,SPT-3G:2022hvq,AtacamaCosmologyTelescope:2025blo}, concerning the amplitude of matter fluctuations inferred from weak lensing or galaxy clustering versus CMB predictions.  Recent improvements in survey systematics and cross-comparisons (e.g.~KiDS, DES, DESI) have reduced, but not eliminated, the tension~\cite{Karim:2024luk,DES:2026fyc,Pantos:2026koc}. These tensions have motivated further scrutiny of alternative models, including those featuring evolving dark energy (DE) or modified gravity (MG)~\cite{CosmoVerseNetwork:2025alb}.
Complementing these tensions, recent observational results provide increasingly strong indications that DE may not be a simple cosmological constant. In particular, DESI, when combined with other probes, such as supernovae, baryon acoustic oscillations (BAO), the CMB, and weak lensing, has found hints that the DE equation of state (EoS) may evolve with time, rather than remaining fixed at $w = -1$~\cite{DESI:2025zgx}. These results, while not yet reaching definitive statistical significance, have strengthened in recent data releases. For example, the significance of evolving DE becomes more pronounced (on the order of $2.8$ to $4.2\sigma$) depending on the combination of datasets used, suggesting that the evidence is no longer marginal~\cite{DESI:2025zgx}. At the same time, other analyses~\cite{Efstathiou:2024xcq,Colgain:2025fct,Huang:2025som,Gialamas:2024lyw} have questioned to what extent these hints are driven by dataset choices or systematic uncertainties. At the same time, MG theories continue to offer viable alternatives to $\Lambda$CDM, capable of modifying gravitational dynamics at large scales or in low density environments~\cite{CosmoVerseNetwork:2025alb}.

Cosmic voids, large underdense regions within the Universe’s LSS~\cite{Gregory:1978qwe,Joeveer:1978ers,vandeWeygaert:2014mqv}, have become powerful cosmological probes~\cite{Pan:2012xxx,Sutter:2012rts,Brouwer:2018xnj,Euclid:2023eom,Pisani:2019cvo,Moresco:2022phi,Fraser:2024ecp,Contarini:2022nvd,Woodfinden:2022bhx,Hamaus:2016wka,Mao:2016faj,Achitouv:2016mbn,Hawken:2016qcy,Hamaus:2017dwj,Achitouv:2019xto,Nadathur:2019mct,Hamaus:2020cbu,eBOSS:2020nuf,Hawken:2019rpp,eBOSS:2020yxq,Woodfinden:2023oca,Contarini:2022mtu,Song:2025vjz} thanks to the advent of extensive, deep redshift surveys~\cite{EUCLID:2011zbd,DESI:2016fyo,LSSTScience:2009jmu} (see~\cite{Contarini:2026yfv} for a review). Voids can be measured
in  many independent ways, yielding a variety of void statistics as cosmological probes~\cite{Alcock:1979mp,Sutter:2012qwe,Lavaux:2011yh,Paz:2013sza,Pisani:2013yxa,Hamaus:2014afa,Hamaus:2015yza,Correa:2021wqw,Amendola:1998xu,Melchior:2013gxd,Clampitt:2014gpa,Chuang:2016wqb,Chantavat:2017ysr,DES:2019zaf,Davies:2020udw,Euclid:2022hdx,Khoraminezhad:2021bdl,Granett:2008ju,Nadathur:2012ksa,Flender:2012wu,Ilic:2013cn,Cai:2013ik,Kovacs:2015bda,Kovacs:2017hxj,DES:2018nlb,Planck:2015fcm,Dong:2020fqt,Hang:2021kfx,Kovacs:2021mnf}, which enable us to
 constrain a diverse set of processes, including DE ~\cite{Lee:2007kq,Lavaux:2009wm,Bos:2012wq,Pisani:2015jha,Verza:2019tvg,Biswas:2010ey},  the effects of neutrino masses~\cite{Massara:2015msa,Banerjee:2016zaa,Sahlen:2018cku,Kreisch:2018var,Kreisch:2021xzq,Schuster:2019hyl,Zhang:2019wtu,Bayer:2021iyb,Verza:2022qsh,Vielzeuf:2023fqw}, dark matter~\cite{Leclercq:2014pga,Yang:2014upa,Reed:2014cta,Baldi:2016oce,Arcari:2022zul}, and, in particular, MG scenarios~\cite{Perico:2019obq,Clampitt:2012ub,Li:2011qda,Spolyar:2013maa,Pollina:2015uaa,Zivick:2014uva,Achitouv:2016jjj,Sahlen:2015wpc,Barreira:2015vra,Cai:2014fma,Voivodic:2016kog,Davies:2019yif,Paillas:2018wxs,Contarini:2020fdu,Mauland:2023eax,Wilson:2022ets,Maggiore:2025mbp}. Indeed,
 screening mechanisms~\cite{Joyce:2014kja} (e.g. chameleon, symmetron, or Vainshtein) in many MG theories become less effective in underdense regions, and voids can amplify the effects of modified forces, offering potentially strong discriminants between GR and its alternatives~\cite{Falck:2017rvl,Voivodic:2016kog,Perico:2019obq,Clampitt:2012ub}.
Previous works have already studied voids in MG scenarios. Examples include theoretical and simulation-based analyses of void abundance~\cite{Clampitt:2012ub,Lam:2014kua,Voivodic:2016kog}, density and velocity profiles~\cite{Cai:2014oea,Perico:2019obq}, tracer bias and environment effects~\citep{Cautun:2017tkc,Perico:2019obq}, and lensing or weak-lensing voids~\citep{Cai:2014oea,Davies:2019yif,Maggiore:2025mbp} in $f(R)$, symmetron, and related MG models. These studies show that under appropriate coupling strengths or unscreened regimes, void interiors become emptier, their compensation walls steeper, and velocity flows around them altered relative to $\Lambda$CDM~\citep{Cai:2014oea,Perico:2019obq,Cataldi:2022lxt}.
However, several challenges remain: many studies neglect baryonic physics or hydrodynamical gas effects; biases arise from the use of tracers (galaxies, halos) rather than the underlying matter field; screening transitions can be complex and environment-dependent; non-linear evolution complicates interpretation; and degeneracies persist with other cosmological effects such as massive neutrinos, redshift-space distortions, and survey systematics~\citep{Cautun:2017tkc,Perico:2019obq,Contarini:2020fdu,Wilson:2022ets}.

The evolution of cosmic voids has traditionally been modeled through  the spherical collapse framework, first developed for overdense halos~\cite{Gunn:1972sv,Peebles:1980yev} and later adapted to describe underdense regions~\cite{Sheth:2003py,Demchenko:2016uzr,Massara:2018dqb}. Within an Einstein de Sitter (EdS) background, this approach yields analytical solutions and a well defined shell-crossing threshold that characterizes the formation of non-linear voids. However, in more realistic cosmological settings such as $\Lambda$CDM or evolving DE models, this condition becomes less tractable and must be determined numerically.
In our earlier work~\cite{Moretti:2025gbp}, we developed a hydrodynamic framework to model the evolution of spherical voids in general cosmologies, capturing their non-linear evolution through the fluid equations of motion in an expanding Universe. In this paper, we extend this approach to MG theories, incorporating an additional scalar degree of freedom and screening effects. We also specialize on the inverse top-hat profile for the density field and on the quasi-static (QS) approximation to derive the main physical quantities. This extension enables a consistent, physically motivated first-step description of void growth beyond GR and provides a powerful tool to link void observables with signatures of modified gravitational dynamics. Additionally, a well-known subtlety of screened MG theories is that the scalar sector can become ill-defined in underdense regions, leading to an imaginary and hence unphysical fifth force~\cite{Barreira:2013eea,Winther:2015pta,Baker:2018mnu,Li:2013tda}. In~\cite{Moretti:2026axy}  this issue was addressed by introducing a void-informed consistency criterion that enforces the existence of a real fifth force during void evolution. Concretely, the criterion can be cast as a condition on background functions and yields a redshift dependent  bound on the attainable void depth, providing a viability requirement during structure formation that is complementary to standard stability requirements. In this work, we apply these results for the first time to the evolution of a single void, using the void-informed criterion to determine the physically admissible dynamics and parameter space in MG.

Furthermore, a key ingredient in any predictive theory of void evolution is the mapping between Lagrangian space, the coordinate label that tracks a given fluid element or mass shell from the initial conditions (ICs), and Eulerian space, i.e. the physical position of that element at a given redshift where void observables are defined and measured. This correspondence is essential because the non-linear dynamics determines how an initially underdense patch expands and rearranges matter, while the quantities inferred from data (void radius, density and velocity profiles, lensing signal) are inherently Eulerian. In this work, we provide this Lagrangian to Eulerian mapping within a MG framework, consistently accounting for the additional scalar degree of freedom and screening. The proposed hydrodynamical approach is particularly well suited for this purpose: by evolving the hydrodynamical equations for spherical shells in an expanding Universe, it yields a physically transparent and self-consistent connection between ICs and the late time, observable void configuration beyond GR.

By combining motivation from cosmic tensions, recent hints of evolving DE, and the sensitivity of voids to departures from GR, we argue that a hydrodynamical approach to void modeling in MG offers a timely and promising path beyond $\Lambda$CDM.

The paper is organized as follows. In section~\ref{sec:void_hydro}, we introduce the hydrodynamical approach. We review the approach within the GR and DE scenarios and then present its generalization to MG. 
We also summarize the main assumptions and discuss the ICs for solving the hydrodynamical system of equations. In section~\ref{sec:Modified_gravity_model_framework}, we introduce the MG framework we adopt and the corresponding equations. We present the concrete model we used to show our main results. We also discuss the viable parameter space in light of the void-informed criteria. In section~\ref{sec:void_evolution}, we present the main results of this work on single void evolution. We assess the impact of MG and discuss void evolution when the pathological branch is considered. We also provide the map from Lagrangian to Eulerian space. Additionally, we discuss shell-crossing. Finally, we conclude in section~\ref{sec:conclusion}.

We close the introduction with table~\ref{tab:definitions} that collects the fundamental definitions that will be used throughout the paper.
\begin{table}[t]
\centering
{
\begin{tabular}{l p{9cm} l l}
\toprule
Symbol & Definition & Space & Dependence  \\
\midrule
$\Delta_{\rm E}$ & 
Mean non-linear matter density contrast & 
Eulerian & $(R_{\rm E},t)$ \\

$\Delta_\mathrm{L}$ & 
Mean linear matter density contrast & 
Lagrangian & $(R_\mathrm{L},t)$   \\

$\delta_{\rm E}$ & 
Non-linear matter density contrast & 
Eulerian & $(\mathbf{r},t)$  \\

$\delta_\mathrm{L}$ & 
Linear matter density contrast & 
Lagrangian & $(\mathbf{r},t)$  \\

$\delta_{\rm v}$ & 
Mapped linear matter density contrast from $\delta_{\rm E}$ at redshift $z$ & 
Lagrangian & $(z,\delta_{\rm E})$  \\

$\delta_{\rm E,sc}$ & 
Non-linear matter density contrast at shell-crossing for a top-hat configuration& Eulerian & (z) \\

$\delta_{\rm v,sc}$ & 
Mapped linear matter density contrast at shell-crossing for a top-hat configuration& 
Lagrangian & $(z)$  \\
\bottomrule
\end{tabular}
\caption{Table of the notation adopted in this paper.}
\label{tab:definitions}
}
\end{table}

\section{The hydrodynamical approach}
\label{sec:void_hydro}

In~\cite{Moretti:2025gbp}, an alternative description for the evolution of isolated, spherically symmetric cosmic voids embedded in a homogeneous and isotropic background has been proposed. It is formulated directly in terms of Newtonian hydrodynamics and uses the matter density contrast, $\delta_{\rm E}$, as the fundamental dynamical variable. Although this strategy is standard in the spherical collapse treatment of overdensities and halo formation~\cite{Padmanabhan:1996qwe,Abramo:2007iu,Pace:2010sn,Pace:2017qxv}, in~\cite{Moretti:2025gbp} we showed that the same set of equations applies to underdensities, with the dynamics differing only through the sign and magnitude of the density contrast, and it can be extended to dynamical DE models. In this work, we extend the formalism to MG. Hereafter, $\rm E$ and $\rm L$ label Eulerian fully non-linear and Lagrangian linear quantities, respectively (see~\cite{Moretti:2025gbp} for a general discussion).

We work in the Newtonian gauge for scalar perturbations of a spatially flat  Friedmann-Lemaître-Robertson-Walker  (FLRW) metric,
\begin{align}
    \mathrm{d}s^2 \,=\,
    -\left[1+2\Psi(\mathbf{x},t)\right]\mathrm{d}t^2
    +a^2(t)\left[1-2\Phi(\mathbf{x},t)\right]\delta_{ij}\mathrm{d}x^i\mathrm{d}x^j\,,
    \label{eq:metric}
\end{align}
where $t$ is the cosmic time, $a(t)$ is the scale factor, $\mathbf{x}$ denotes the comoving spatial coordinate vector, $\delta_{ij}$ is the three-dimensional Kronecker symbol, and $\Phi$ and $\Psi$ are the gravitational potentials. We assume spherical symmetry, retain only pressureless matter and dark energy, and treat the latter as a smooth background component. The matter stress-energy tensor is taken to be that of a pressureless perfect fluid, $
    T_{\mu\nu} \,=\, \rho_{\rm m}\,u_\mu u_\nu\,$,
with $\rho_{\rm m}$ the matter density and $u_\mu$ the four-velocity.

\subsection{GR limit and dynamical dark energy}

Assuming GR and a background that may include dynamical DE, in the Newtonian (sub-horizon, weak-field) regime, the perturbed Einstein equations yield the usual Poisson relation for $\Psi$, together with the lensing combination sourced by the same density perturbation, i.e.
\begin{align}
    \nabla^2_{\mathbf{x}}\Psi &\,=\, 4\pi G a^2 \bar{\rho}_{\rm m}\,\delta_{\rm E}\,,
    \label{eq:poisson_GR_rewrite}\\
    \nabla^2_{\mathbf{x}}\left(\Phi+\Psi\right) &\,=\, 8\pi G a^2 \bar{\rho}_{\rm m}\,\delta_{\rm E}\,,
    \label{eq:lensing_GR_rewrite}
\end{align}
where $\bar{\rho}_{\mathrm{m}}(t)$ is the background matter density and $G$ is the Newtonian gravitational constant. For dust, the absence of anisotropic stress implies $\Phi=\Psi$~\cite{Kodama:1984ziu}. Conservation of $T_{\mu\nu}$ provides the continuity and Euler equations for the comoving peculiar velocity field $\vec{u}$~\cite{Padmanabhan:1996qwe}, i.e.
\begin{align}
    \frac{\partial\delta_{\rm E}}{\partial t} + (1+\delta_{\rm E})\,\nabla_{\mathbf{x}}\!\cdot\vec{u} &= 0,
    \label{eq:cont_rewrite}\\
    \frac{\partial\vec{u}}{\partial t} + 2H\vec{u} + (\vec{u}\cdot\nabla_{\mathbf{x}})\vec{u} + \frac{1}{a^2}\nabla_{\mathbf{x}}\Psi &= 0,
    \label{eq:euler_rewrite}
\end{align}
where $H\equiv ({\rm d} a/{\rm d}t)/a$. Combining eqs.~\eqref{eq:poisson_GR_rewrite}--\eqref{eq:euler_rewrite} under spherical symmetry leads to the familiar non-linear evolution equation for the Eulerian density contrast~\cite{Padmanabhan:1996qwe},
\begin{align}
    \delta_{\rm E}'' + \left(2+\frac{H'}{H}\right)\delta_{\rm E}'
    -\frac{4}{3}\frac{(\delta_{\rm E}')^2}{1+\delta_{\rm E}}
    -\frac{3}{2}\Omega_{\rm m}\,(1+\delta_{\rm E})\,\delta_{\rm E}
    =0\,,
    \label{eq:delta_nl_GR_rewrite}
\end{align}
where primes denote $\partial/\partial \ln{a}$, and $\Omega_{\rm m}=\bar\rho_{\rm m}/3M_{\rm pl}^2H^2$ is the background matter fraction  and $M_{\rm pl}^2=(8\pi G)^{-1}$. Linearizing eq.~\eqref{eq:delta_nl_GR_rewrite} gives
\begin{align}
    \delta_{\rm L}'' + \left(2+\frac{H'}{H}\right)\delta_{\rm L}'
    -\frac{3}{2}\Omega_{\rm m}\,\delta_{\rm L}
    =0\,.
    \label{eq:delta_lin_GR_rewrite}
\end{align}
The spatial dependence enters only through the initial profile and, hence, remains parametric in this formulation.

\subsection{MG generalization}
\label{eq:MG_generalization}

For a broad class of MG models with minimally coupled matter, the fluid equations~\eqref{eq:cont_rewrite} and~\eqref{eq:euler_rewrite} retain their form, while the relationship between the potentials and the matter perturbation is modified. A convenient and widely used way to encode these changes is via two functions, $\mu_{\rm NL}$ and $\Sigma_{\rm NL}$, that generalize the Poisson and lensing equations~\cite{Zhang:2007nk,Amendola:2007rr,Pogosian:2010tj}, which here we also employ at non-linear scale~\footnote{Let us note that if the gravitational couplings in eqs.~\eqref{eq:poisson_MG_rewrite} and~\eqref{eq:lensing_MG_rewrite} depend on the spatial coordinate, then these relations are not equivalent to their Fourier-space counterparts.}:
\begin{align}
    \nabla^2_{\mathbf{x}}\Psi &\,=\, 4\pi G a^2 \bar{\rho}_{\rm m}\,\,\mu_{\rm NL}(a,|\mathbf{x}|)\,\delta_{\rm E}\,,
    \label{eq:poisson_MG_rewrite}\\
    \nabla^2_{\mathbf{x}}\left(\Phi+\Psi\right) &\,=\, 8\pi G a^2 \bar{\rho}_{\rm m}\,\Sigma_{\rm NL}(a,|\mathbf{x}|)\,\delta_{\rm E}\,,
    \label{eq:lensing_MG_rewrite}
\end{align}
together with the gravitational slip parameter $\eta_{\rm NL}(a,|\mathbf{x}|)\equiv \Phi/\Psi$. In GR, these functions are equal to unity. In screened theories, these modifications are typically scale- and environment-dependent; in a spherical setup, this translates into an effective coupling that can depend on time and on the spherical perturbation configuration (e.g., radius and density contrast).

Repeating the derivation that leads to eq.~\eqref{eq:delta_nl_GR_rewrite}, one finds that the void evolution equation is modified only through the gravitational source term, which becomes proportional to the effective coupling, i.e.
\begin{align}
    \delta_{\rm E}'' + \left(2+\frac{H'}{H}\right)\delta_{\rm E}'
    -\frac{4}{3}\frac{(\delta_{\rm E}')^2}{1+\delta_{\rm E}}
    -\frac{3}{2}\Omega_{\rm m}\,\mu_{\rm NL}(a,R)\,(1+\delta_{\rm E})\,\delta_{\rm E}
    =0\,.
    \label{eq:delta_nl_MG_rewrite}
    \end{align}
     At linear order, the growth equation generalizes to
    \begin{align}
    \delta_{\rm L}'' + \left(2+\frac{H'}{H}\right)\delta_{\rm L}'
    -\frac{3}{2}\Omega_{\rm m}\,\mu_{\rm L}(a,R)\,\delta_{\rm L}
    =0\,,
    \label{eq:delta_lin_MG_rewrite}
\end{align}
where $\mu_{\rm L}$ is the modification to the gravitational coupling at linear scales  (often $\mu_{\rm L}=\mu$).

In the following, the specific form of $\mu_{\rm NL}$ appropriate to the MG scenario under consideration will be provided in the following section, allowing us to connect the non-linear void dynamics to the underlying theory.

\subsection{Setup for isolated void evolution}
\label{Eq:setup_for_isolated_void_evolution}

We now introduce the assumptions we adopt to model the evolution of a single isolated void in a cosmological background, i.e. how we practically solve the equations presented in the previous section. 

The non-linear hydrodynamical equation, eq.~\eqref{eq:delta_nl_GR_rewrite} in GR and eq.~\eqref{eq:delta_nl_MG_rewrite} in MG, is valid for any spherically symmetric matter density profile. Additionally, we specify that the initial void density profile is an inverse top-hat~\cite{Moretti:2025gbp}, defined as
\begin{align}
    \delta_{\rm E}(t_{\rm in}, r_{\rm in}) \,=\,
    \begin{cases}
    \delta_{\rm v,in} & \text{for } r_{\rm in} \leq r_{\rm v,in} \\
    0 & \text{for } r_{\rm in} > r_{\rm v,in}
    \end{cases}\,,
    \label{Eq:initial_top_hat_delta}
\end{align}
where $t_{\rm in}$ denotes the initial time, ${r}_{\rm in}=a_{\rm in}|\mathbf{x}|$, $\delta_{\rm v,in}<0$ is the initial void depth, and $r_{\rm v,in}$ is the initial void radius. Imposing eq.~\eqref{Eq:initial_top_hat_delta} ensures that the top-hat structure is preserved at later times.
This holds because all points with $r_{\rm in}<r_{\rm v,in}$ start from identical ICs and, because the evolution equations are local, they satisfy the same evolution equation, thus implying identical solutions up to shell-crossing. In GR, this is because the only spatial dependence in eq.~\eqref{eq:delta_nl_GR_rewrite} enters parametrically through the initial profile. In MG, the same conclusion holds because the only difference in eq.~\eqref{eq:delta_nl_MG_rewrite} with respect to GR enters through the non-linear gravitational potential, which in principle could introduce an explicit dependence on the void scale, but for the models considered in this work, it depends only on $a$ and $\delta_{\rm E}$.
Possible extensions in which the potentials acquire an explicit radial dependence are beyond the scope of this work.
The same reasoning applies to linear equations, i.e., eqs.~\eqref{eq:delta_lin_GR_rewrite} and~\eqref{eq:delta_lin_MG_rewrite}.

In practice, we solve the evolution equations for a generic (representative) point inside the void. We integrate eqs.~\eqref{eq:delta_nl_MG_rewrite} and~\eqref{eq:delta_lin_MG_rewrite}, which are second order ordinary differential equations and therefore require two ICs. We start the integration at $a_{\rm in}= 10^{-7}$ during matter domination, when MG effects are negligible, and the dynamics is reduced to GR. Since radiation is neglected in our setup, this corresponds to an idealized early-time EdS regime. This choice is not intended to be physically realistic, but is introduced only to fix the ICs; our late-time results are insensitive to this assumption (see appendix~B of~\cite{Moretti:2025gbp}). We adopt two early time assumptions: perturbations are in the linear regime, so that $\delta_{\rm E}=\delta_{\rm L}$, and the decaying mode is neglected. These imply
\begin{align}
    \delta_{\rm E}'(a_{\rm in}) = \delta_{\rm E}(a_{\rm in}) \equiv \delta_{\rm v,in}\,.
    \label{eq:ICs_hydrodynamical_eqs}
\end{align}
The value of $\delta_{\rm v,in}$ is fixed a posteriori by imposing a late time condition on the non-linear density contrast, for instance $\delta_{\rm E}(z=0)=-0.5$, and determining $\delta_{\rm v,in}$ through a shooting procedure. Extended discussions, numerical details, and tests of the impact of the choice of $a_{\rm in}$ and the adopted ICs are presented in~\cite{Moretti:2025gbp}.

The dynamics described above is valid up to shell-crossing, which, under the top-hat assumption, is defined as the time at which the outermost void shell reaches the surrounding environment~\cite{Moretti:2025gbp}. The shell-crossing condition reads
\begin{align}
    \frac{\mathrm{d} \delta_{\rm E} }{\mathrm{d} \delta_{\rm v,in}} \,=\, -\frac{(1+\delta_{\rm E})}{(1+\delta_{\rm v,in})\,\delta_{\rm v,in}} \,.
    \label{Eq:Shell_crossing_Delta}
\end{align}
We refer to~\cite{Moretti:2025gbp} for the derivation and a more detailed discussion.

Before proceeding, two brief comments are useful to clarify our notation and complete the model setup. First, in the literature, the void dynamics is often expressed in terms of the averaged density contrasts
\begin{align}
    \Delta_{\rm E}(R_{\rm E},t) &= \frac{3}{4\pi R_{\rm E}^{3}}\int_0^{R_{\rm E}}{\rm d}s\,s^{2}\int\mathrm{d}\Omega\,\delta_{\rm E}(\mathbf{s},t)\,,\\
   \Delta_{\rm L}(R_{\rm L},t) &= \frac{3}{4\pi R_{\rm L}^{3}}\int_0^{R_{\rm L}}{\rm d}s\,s^{2}\int\mathrm{d}\Omega\,\delta_{\rm L}(\mathbf{s},t)\,,
    \label{eq:mean_density_contrast}
\end{align}
where $s=\lvert\mathbf{s}\rvert$ and $\Omega$ is the solid angle. In a top-hat configuration, $\delta_{\rm E}$ ($\delta_{\rm L}$) is spatially constant within the underdense region, so $\Delta_{\rm E}$ ($\Delta_{\rm L}$) and $\delta_{\rm E}$ ($\delta_{\rm L}$) coincide there. Therefore, using averaged or local quantities leads to the same dynamics for the top-hat voids considered in this work. A more detailed discussion is given in~\cite{Moretti:2025gbp}.

Second, within the hydrodynamical model, the evolution of the void radius can always be reconstructed from the density evolution by enforcing mass conservation for spherical shells~\cite{Moretti:2025gbp}. Denoting $R(t)$ as the physical radius of the outermost shell, the enclosed mass reads
\begin{align}
    M_{\rm v} = \frac{4\pi}{3}\,\bar{\rho}_{\rm m}(t)\,\bigl[1+\Delta_{\rm E}(t)\bigr]\,R^{3}(t)\,,
    \label{eq:mass_conservation}
\end{align}
and is conserved throughout evolution. For a top-hat, one may equivalently replace $\Delta_{\rm E}$ with $\delta_{\rm E}$ in eq.~\eqref{eq:mass_conservation}, so that the radius evolution follows directly once $\delta_{\rm E}(t)$ is known.

\section{Modified gravity model framework}
\label{sec:Modified_gravity_model_framework}

The hydrodynamic framework developed in the previous section is theory-agnostic: it applies to any cosmological model in which matter follows standard conservation equations, while departures from GR enter through the relation between the gravitational potentials and the matter density perturbation. In the remainder of this work, we focus on the class of Galileon models~\cite{Horndeski:1974wa,Deffayet:2009mn,Nicolis:2008in,Deffayet:2009wt,Deffayet:2011gz,DeFelice:2010pv}, for which these modifications are generated by derivative scalar interactions and are regulated on small scales by Vainshtein screening~\cite{Vainshtein:1972sx}. Accordingly, we will specify the corresponding expressions for the effective gravitational coupling (and related functions) that determine the void dynamics in this theory space.

\subsection{A model-independent approach}
\label{sec:EFT}

To provide concrete examples of cosmological models beyond $\Lambda$CDM, we employ the Effective Field Theory (EFT) approach to DE and MG~\cite{Gubitosi:2012hu,Bloomfield:2012ff,Frusciante:2019xia}. The EFT provides a model-independent description of perturbations around a FLRW background, constructed by organizing all operators compatible with the symmetries of the problem in the unitary gauge and tracking their impact on cosmological observables. This framework can be systematically extended towards mildly non-linear scales~\cite{Bellini:2015wfa,Frusciante:2017nfr,Yamauchi:2017ibz,Cusin:2017mzw,Cusin:2017wjg}. Adopting this EFT-based, model-independent description of the gravitational sector, we characterize departures from $\Lambda$CDM directly at the perturbative level through the time dependence of the EFT functions (subject to the viability requirements described in section~\ref{sec:viability}), rather than working with the covariant Galileon functions, $G_i(\phi,X)$ with $\phi$ being the scalar field and $X$ its kinetic term. This choice provides a transparent link between the underlying theory space and the phenomenology probed by LSS, lensing, and GW propagation, while enabling a unified treatment of different realizations within the same EFT framework.

For Galileon theories, it is advantageous to trade the EFT coefficients for a small set of physically interpretable functions of time. A widely used parametrization is the \emph{$\alpha$-basis} introduced in~\cite{Bellini:2014fua,Bellini:2015wfa}, in which the dynamics of perturbations is fully specified by the dimensionless functions
$
\{\alpha_{\rm K},\alpha_{\rm B},\alpha_{\rm M},\alpha_{\rm T},\alpha_{\rm V1},\alpha_{\rm V2},\alpha_{\rm V3}\}$, 
together with the background expansion history.

In this language, $\alpha_{\rm K}$ controls the kinetic energy of scalar fluctuations and contributes to determining their effective sound speed. However, it has been shown that it cannot be bounded by data~\cite{Frusciante:2018jzw}, and that fixing its value does not impact the constraints~\cite{Bellini:2015xja}. The braiding parameter $\alpha_{\rm B}$ quantifies kinetic mixing between the scalar and metric sectors, a characteristic source of scale-dependent modifications to clustering and lensing. Variations of the effective gravitational coupling are encoded by the Planck-mass running,
\begin{align}
    \alpha_{\rm M} \equiv \frac{{\rm d}\ln M^2}{{\rm d}\ln a}\,,
    \label{eq:alpha_M_running_planck_mass}
\end{align}
with $M^2(t)$ as the effective Planck mass. Moreover, deviations from luminal propagation of tensor modes are described by the tensor-speed excess $\alpha_{\rm T}$ through
\begin{align}
    c_{\rm T}^2 = 1+\alpha_{\rm T}\,,
\end{align}
which is tightly constrained by multimessenger observations, notably GW170817 and its electromagnetic counterpart, to be extremely close to zero, i.e. $\alpha_{\rm T}\sim O(10^{-15})$~\cite{LIGOScientific:2017zic}. Finally, the $\alpha_{{\rm V}i}$-functions define dimensionless time-dependent functions parameterizing the action purely at non-linear scales~\cite{Bellini:2015wfa,Frusciante:2017nfr,Cusin:2017mzw,Yamauchi:2017ibz}.

We restrict our analysis to the class of models in which GWs propagate at the speed of light, as required by current constraints. Accordingly, throughout this work, we set
\begin{align}
    \alpha_{\rm T} = 0, \qquad \alpha_{V_i}=0 \ \ (i=1,2,3)\,.
\end{align}

\subsection{Modified gravitational potentials and screening scale}
\label{sec:Modified_gravitational_potentials_and_screening_scale}

In this section, we quantify how the gravitational potentials are modified around spherical voids when non-linearities in the scalar
sector are retained. We work on sub-horizon scales and adopt the quasi-static approximation (QSA), in which time derivatives of the
perturbations are neglected with respect to spatial gradients. Within this regime, the dynamics of non-relativistic tracers can be cast in terms of an effective Poisson relation for $\Psi$.
Our goal is to derive a compact expression for the corresponding non-linear modification function, $\mu_{\rm NL}$, and to make explicit the role of derivative self-interactions (Vainshtein-type screening) in void environments.

We consider the non-linear system of equations derived in~\cite{Kimura:2011dc} and analyze it on sub-horizon scales while applying the QSA. Then, the metric potentials and the scalar fluctuation $\chi\equiv \delta\phi/\dot{\phi}$, where $\phi$ is the scalar field and a dot denotes a time derivative,  obey the following equations \footnote{For $\alpha_{\rm B}$ we use the definition in~\cite{Bellini:2014fua}.}:
\begin{align}
    0\,=\,&\nabla^2\!\left(\Phi-\Psi-\alpha_{\rm M}H\chi\right)\,,
    \label{eq:shift_void}\\
    0\,=\,&M^2\nabla^2\Phi-\frac{1}{2}\left(a^2\bar\rho_{\rm m}\,\delta_{\rm E}-M^2\alpha_{\rm B} H\nabla^2\chi\right)\,,
    \label{eq:PoissonPhi_void}\\
    0\,=\,&-\frac{H}{2}\!\left[\alpha c_{\rm s}^2+\frac{\alpha_{\rm B}^2}{2}+2\alpha_{\rm B}\alpha_{\rm M}\right]\!\nabla^2\chi
    -\alpha_{\rm M}\nabla^2\Phi-\frac{1}{2}\alpha_{\rm B}\nabla^2\Psi
    +\frac{1}{2a^2}(\alpha_{\rm B}+\alpha_{\rm M})\chi^{(2)}\,,
    \label{eq:scalar2_void}
\end{align}
where $ \chi^{(2)}\equiv(\nabla^2\chi)^2-(\nabla_i\nabla_j\chi)^2$ and 
\begin{align}
    \alpha c_{\rm s}^{2}\,=\, -(2 - \alpha_{\rm B})\left(\frac{H'}{H} - \frac{\alpha_{\rm B}}{2} - \alpha_{\rm M} \right)
    + \alpha'_{\rm B} - \frac{\bar{\rho}_{\rm m}}{M^2}\,,
\end{align}
being $c_{\rm s}$ the speed of propagation of the scalar mode. Combining the above relations, one can eliminate $\Phi$ and $\Psi$ and obtain a single  equation for the scalar mode
\begin{align}
    \nabla^2\chi+\frac{\lambda^2}{a^2}\Big[(\nabla_i\nabla_j\chi)^2-(\nabla^2\chi)^2\Big]
    =-4\pi G\,a^2\,\beta^2\,\bar\rho_{\rm m}\,\delta_{\rm E}\,,
    \label{eq:chi_master_void_merged}
\end{align}
where the parameters controlling the strength of the scalar coupling and the derivative self-interactions are
\begin{align}
    \lambda^2\equiv\frac{\alpha_{\rm B}+\alpha_{\rm M}}{H\,\alpha c_{\rm s}^2},
    \qquad
    \beta^2\equiv\frac{2\!\left(\alpha_{\rm M}+\frac{1}{2}\alpha_{\rm B}\right)}{H\,\alpha c_{\rm s}^2} \frac{M_{\rm pl}^2}{M^2}\,.
\label{eq:lambda_beta_void_merged}
\end{align}
The non-linear term proportional to $\lambda^2$ is responsible for Vainshtein-type screening. 

Assuming spherical symmetry, $\chi=\chi(r)$, eq.~\eqref{eq:chi_master_void_merged} can be integrated once and expressed in terms
of the enclosed defect mass contrast
\begin{align}
    m(r)\equiv 4\pi\!\int_0^r\!dr'\,r'^2\,\bar\rho_{\rm m}\,\delta_{\rm E}(r')\,,
\end{align}
which yields an algebraic (quadratic) relation for the scalar gradient. We note that with this definition, $m<0$. It is convenient to introduce a Vainshtein scale 
\begin{align}
    r_{\rm V}^3(r)\equiv 8G\,\beta^2\lambda^2\,m(r)\,,
    \label{eq:rV_void_merged}
\end{align}
so that the solution  can be written schematically as
\begin{align}
    \chi'(r)\ \propto\ \frac{r a^2}{\lambda^2}\left[1-\sqrt{1+\frac{r_{\rm V}^3(r)}{r^3}}\right]\,.
\end{align}
It follows from the definition of $r_{\rm V}$ that the Vainshtein scale is negative. This is common in literature~\cite{Barreira:2015vra}.
For an inverse top-hat void of radius $R$ and constant $\delta_{\rm E}<0$, the enclosed defect mass is $m(R)=\frac{4\pi}{3}\bar\rho_{\rm m}R^3 \delta_{\rm E} $, and the corresponding Vainshtein scale reads 
\begin{align}
    \frac{R_{\rm V}^3}{R^3} = \frac{32 \pi}{3} G\,\beta^2\lambda^2 \bar\rho_{\rm m}\,\delta_{\rm E}\,.
    \label{eq:RV_tophat_void_merged}
\end{align}
Substituting into the QSA relations for the potentials yields a modified Poisson equation of the form~\eqref{eq:poisson_MG_rewrite}
with
\begin{align}
    \mu_{\rm NL}(a,R) &\,=\, \frac{M_{\rm pl}^2}{M^{2}}\Bigg\{1
    + 2\left(\frac{M^2}{M_{\rm pl}^{2}}\mu_{\rm L} - 1\right)
    \left(\frac{R}{R_{\rm V}}\right)^{3}
    \bigg[\sqrt{1+\left(\frac{R_{\rm V}}{R}\right)^{3}}- 1\bigg]\Bigg\}\,,
    \label{eq:non_linear_gravitational_potential}
\end{align}
where 
\begin{align}
    \mu_{\rm L}(a) \,=\, \frac{M_{\rm pl}^2}{M^{2}}\left[1 + \frac{2\left(\frac{1}{2}\alpha_{\rm B} + \alpha_{\rm M}\right)^{2}}{\alpha c_{\rm s}^{2}}\right]\,.
    \label{eq:linear_potential}
\end{align}

\subsection{A concrete MG model}
\label{sec:a_concrete_MG_model}

We now specify the time dependence of the $\alpha$-functions and the expansion history adopted in this work. The background expansion is parameterized through the effective DE EoS for which we assume the CPL form~\cite{Chevallier:2000qy,Linder:2002et}
\begin{align}
    w_{\rm DE}(a)=w_0+w_a(1-a)\,,
    \label{eq:CPL}
\end{align}
with constant parameters $(w_0,w_a)$. This choice uniquely determines $H(a)$ and $\Omega_{\rm DE}(a)$.

Deviations from $\Lambda$CDM at the level relevant for the void dynamics are then encoded in the braiding function, which we model with the two-parameter ansatz proposed in~\cite{Traykova:2021hbr},
\begin{align}
    \alpha_{\rm B}(a)=\alpha_{\rm B_0}\left(\frac{H_0}{H(a)}\right)^{\frac{4}{m}}\,,
    \label{eq:alphaB_benchmark}
\end{align}
where $\alpha_{\rm B_0}$ sets the amplitude and $m$ controls the redshift evolution. We fix the effective Planck mass to its standard value, $M=M_{\rm pl}$, and therefore $\alpha_{\rm M}=0$. The functional form of $\alpha_{\rm K}$ is not relevant because its contribution to observables has been shown to be below the cosmic variance~\cite{Frusciante:2018jzw}. Therefore, it is usually kept to a fixed value for stability purposes, and the chosen value does not affect constraints of the data on other model parameters~\cite{Bellini:2015xja}. We also note that $\alpha_{\rm K}$ does not enter any of the expressions relevant for our analysis. The model is therefore specified by four parameters, $\{\alpha_{\rm B_0},\, m,\, w_0,\, w_a\}\,$. This parametrization and the CPL EoS for the background describe shift-symmetric scalar-tensor theories.  

Existing analyses already place informative bounds on the model’s free parameters, using CMB temperature/polarization auto- and cross-correlations and the lensing potential, complemented by BAO, RSD, and SNIa measurements~\cite{Traykova:2021hbr}.
Euclid forecasts likewise show that $\alpha_{\rm B_0}$ tightens substantially with Euclid alone, whereas improvements on $m$ hinge on combining spectroscopic galaxy clustering  with the photometric $3\times 2$pt statistics~\cite{Euclid:2025tpw}. On smaller scales, spherical-collapse and halo mass-function calculations suggest an enhanced abundance of massive haloes relative to $\Lambda$CDM for masses larger than $10^{14}\,h^{-1}M_\odot$~\cite{Albuquerque:2024hwv} and the implications for cosmic-void abundances have also been investigated recently~\cite{Takadera:2025ehm}.

We adopt a \textit{baseline} background expansion history defined by the CPL EoS parameters and the present day matter abundance,
\begin{align}
    w_0=-0.97,\qquad w_a=-0.11,\qquad \Omega_{{\rm m},0}=0.32\,.
    \label{eq:baseline_background}
\end{align}
Since we set $\Omega_{{\rm r},0}=0$ and assume spatial flatness, the present day DE abundance follows immediately from $\Omega_{{\rm m},0}$.
The braiding amplitude, $\alpha_{\rm B_0}$, and the  parameter controlling the time evolution, $m$, in eq.~\eqref{eq:alphaB_benchmark} will be varied and their values will be chosen to visualize and quantify the modifications. 

\subsection{Viable parameter space}
\label{sec:viability}

In exploring the parameter space, we retain only models that meet a set of theoretical and observational viability requirements.
First, we impose \emph{theoretical stability} of scalar perturbations by enforcing the absence of ghosts and gradient instabilities throughout the redshift range relevant for structure formation. In the EFT framework, this amounts to requiring that, at all redshifts considered,
\begin{align}
    \alpha \equiv \alpha_{\rm K} + \frac{3}{2}\,\alpha_{\rm B}^2 > 0\,,
    \qquad 
    c_{\rm s}^2>0\,.
\end{align}
Second, as already discussed, we enforce the multimessenger bound on the speed of GWs by imposing luminal tensor propagation, i.e. $\alpha_{\rm T} = 0$, at all redshifts.

Finally, we apply the void-informed theoretical bound of~\cite{Moretti:2026axy}, which ensures that the non-linear effective gravitational coupling inside spherical voids remains real, i.e. avoids the imaginary branch pathology of the fifth force, which has been identified in several works~\cite{Barreira:2013eea,Barreira:2015vra,Winther:2015pta,Takadera:2025ehm,Baker:2018mnu}. In this framework, the key background function $f_{\rm MG}(a)$ is defined through the ratio of the Vainshtein and physical radii,
\begin{align}
    \left(\frac{R_{\rm V}}{R}\right)^3 \equiv f_{\rm MG}(a)\,\delta_{\rm E}\,, \quad\text{with}\quad\
    f_{\rm MG}(a)= \frac{32 \pi}{3} G\,\beta^2\lambda^2 \bar\rho_{\rm m}\,.\label{eq:fMG}
\end{align}
Requiring the square root argument in the non-linear coupling to remain non-negative for all physical void configurations ($\delta_{\rm E}\ge -1$)  implies
\begin{align}
    1+f_{\rm MG}(a)\,\delta_{\rm E} \ge 0\,.
    \label{eq:square_argument_stability}
\end{align}
Since deep voids can approach $\delta_{\rm E}= -1$, we enforce the conservative viability criterion introduced in~\cite{Moretti:2026axy}
\begin{align}
    \max_{0\le z\le z_{\rm in}} f_{\rm MG}(z) > 1 \,,
    \qquad \text{with }\,z_{\rm in}\,=\,100\,, \qquad \Rightarrow \quad \text{model excluded}.
    \label{Eq:viability_criterion_voids_MG}
\end{align}
Equivalently, eq.~\eqref{eq:square_argument_stability} yields a redshift-dependent minimum allowed void depth\cite{Moretti:2026axy}
\begin{align}
    \delta_{\min}(z)=\max\!\left[-1,\,-\frac{1}{f_{\rm MG}(z)}\right]\,.
    \label{eq:minimum_allowed_void_depth}
\end{align}

We now apply the viability requirements discussed above to the model introduced in section~\ref{sec:a_concrete_MG_model}. For the MG sector, we scan the $1\sigma$ region in the $(\alpha_{\rm B_0},m)$ plane reported in Table I of~\cite{Traykova:2021hbr} for the combined data set and the $\Lambda=0$ branch, i.e. $\alpha_{\rm B_0}\in[0.3,0.9]$ and $m\in[2.0,2.8]$. These parameter ranges already satisfy theoretical stability requirements. We first determine the viable subset of the scanned parameter space and then quantify how the bound translates into a redshift-dependent minimum allowed void depth $\delta_{\min}(z)$.

\begin{figure}[t]
    \centering
    \includegraphics[width=1.0\linewidth]{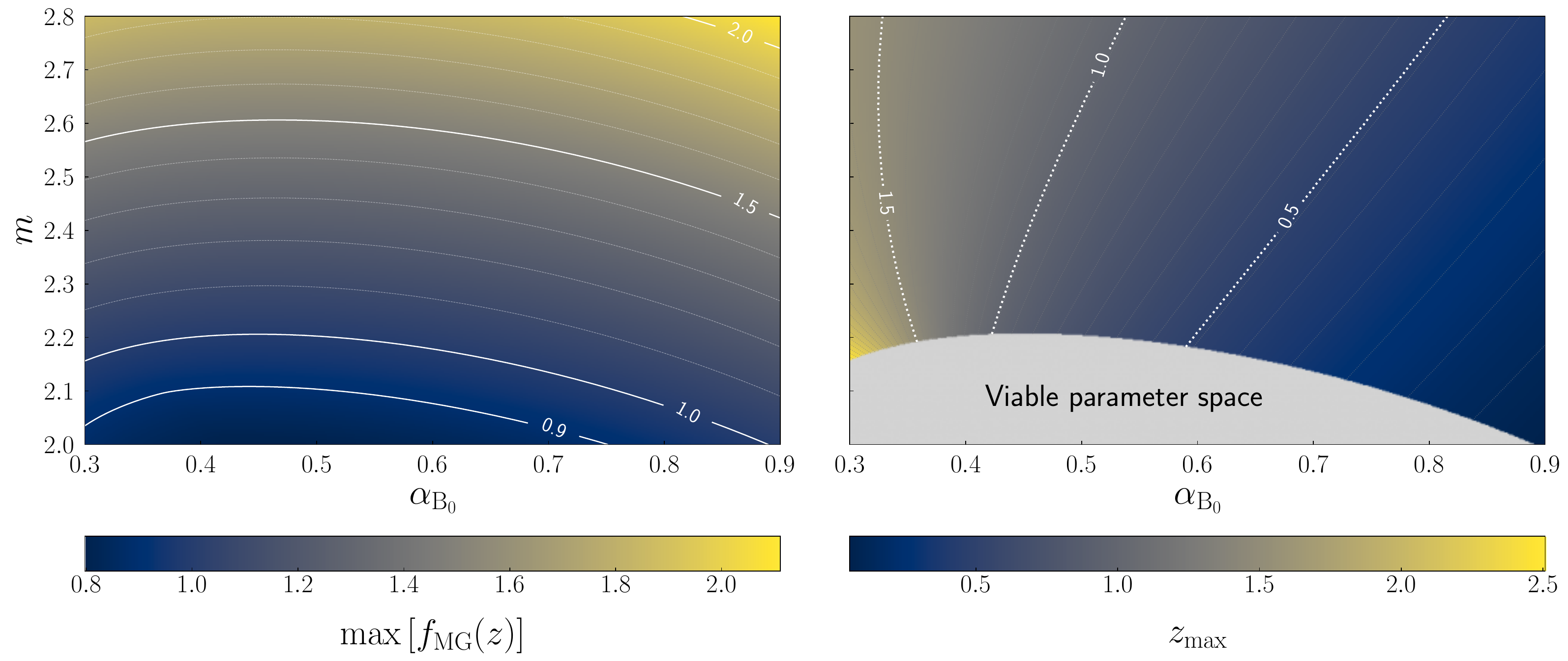}
    \caption{Void-informed viability across the $(\alpha_{\rm B_0},m)$ plane for the baseline background parameters given in eq.~\eqref{eq:baseline_background}. The scanned ranges correspond to the $1\sigma$ combined constraints of~\cite{Traykova:2021hbr} for the $\Lambda=0$ branch.
    \textbf{Left:} maximum value attained by $f_{\rm MG}(z)$ over $0\le z\le 100$ for each parameter point, with contours indicating constant values of this maximum. 
    \textbf{Right:} models satisfying the void-informed bound in eq.~\eqref{Eq:viability_criterion_voids_MG}, shown as the gray region. Outside the gray region, models are excluded and are color coded by the redshift $z_{\rm max}$ at which $f_{\rm MG}(z)$ reaches its maximum.}
    \label{fig:max_funz_sqrt}
\end{figure}

Figure~\ref{fig:max_funz_sqrt} shows the results of enforcing the void-informed viability requirement in eq.~\eqref{Eq:viability_criterion_voids_MG} in the $(\alpha_{\rm B_0},m)$ plane. In the left panel, each point is assigned the maximum value achieved by $f_{\rm MG}(z)$ over $0\le z\le 100$, and the contour lines mark constant values of this maximum across the plane. The right panel identifies the subset of models that satisfy the bound, shown as the gray region, where the peak value remains below unity. Models that fail the test are discarded and are color coded according to the redshift $z_{\rm max}$ at which the maximum occurs. Among the scanned parameters, the filter excludes most of the parameter points, leaving only about $18\%$ viable ones, with maxima typically occurring at $z_{\rm max}\lesssim 1.5$, i.e., during the epoch of structure formation. Compared to the parametrization adopted in~\cite{Moretti:2026axy}, the present one strengthens the impact of the void-informed condition, removing a larger fraction of the parameter space favored by current data.

\begin{figure}[t]
    \centering
    \includegraphics[width=1.0\linewidth]{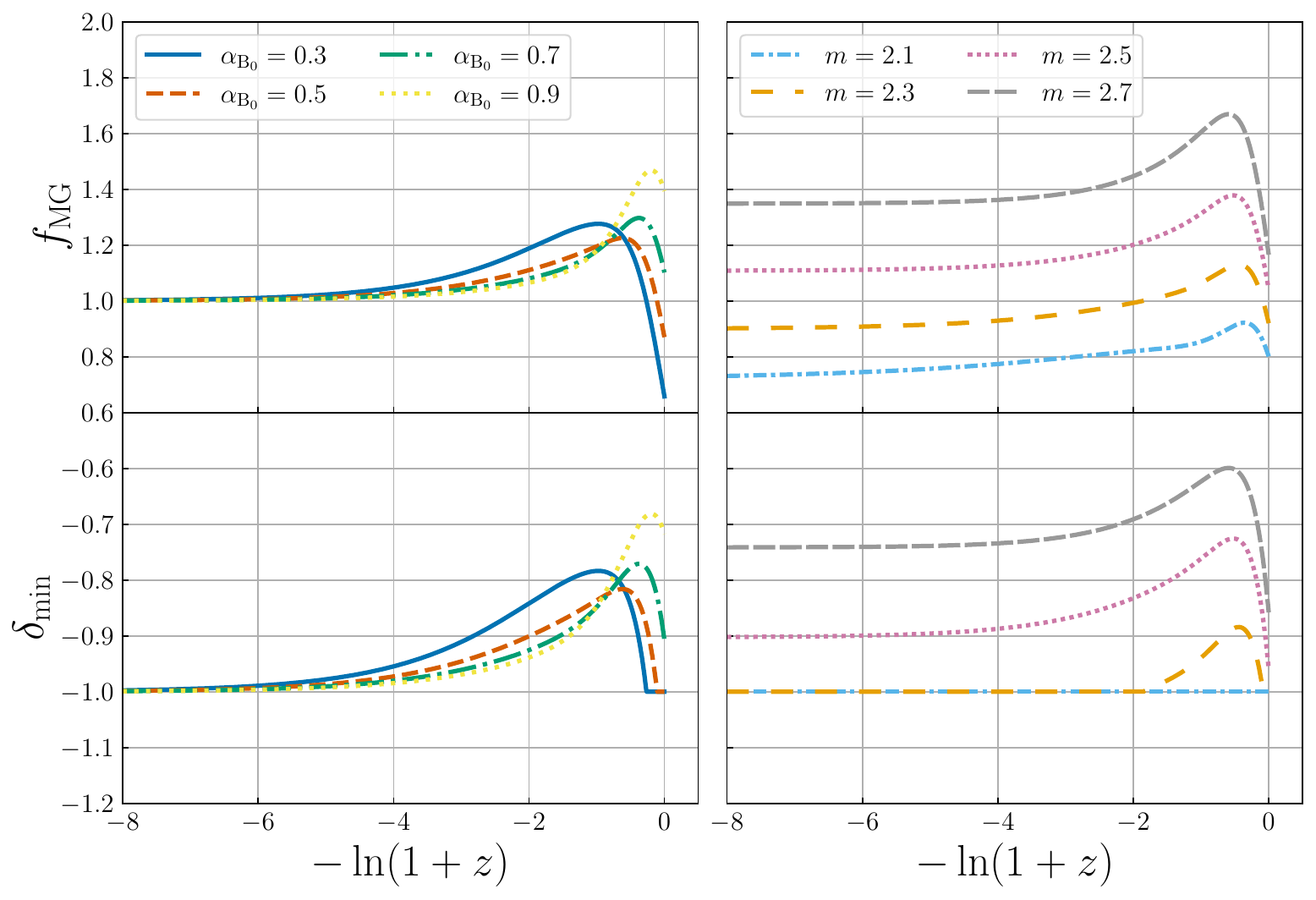}
    \caption{Evolution of $f_{\rm MG}(z)$ and $\delta_{\min}(z)$ as a function of redshift (top and bottom rows, respectively). The left column varies $\alpha_{\rm B_0}\in\{0.3,0.5,0.7,0.9\}$ at fixed $m$, while the right column varies $m\in\{2.1,2.3,2.5,2.7\}$ at fixed $\alpha_{\rm B_0}$. In each panel, the remaining parameters are set to the best fit of~\cite{Traykova:2021hbr}, $\alpha_{\rm B_0}=0.6$ and $m=2.4$.
    }
\label{fig:f_MG_delta_min}
\end{figure}
We now investigate the redshift dependent minimum allowed void depth in eq.~\eqref{eq:minimum_allowed_void_depth}.
To this end, we focus on the best fit point of~\cite{Traykova:2021hbr}, $\alpha_{\rm B_0}=0.6$ and $m=2.4$, and vary one MG parameter at a time while keeping the other fixed. The results are shown in figure~\ref{fig:f_MG_delta_min}, which is organized into four panels. The top row displays the evolution of $f_{\rm MG}$ as a function of $-\ln(1+z)\in[-8,0]$, while the bottom row reports the corresponding $\delta_{\min}$. The left column varies $\alpha_{\rm B_0}$ at fixed $m$, taking $\alpha_{\rm B_0}\in\{0.3,0.5,0.7,0.9\}$, while the right column varies $m$ at fixed $\alpha_{\rm B_0}$, with $m\in\{2.1,2.3,2.5,2.7\}$.
Two features are immediate. First, the region of $f_{\rm MG}(z)>1$ is not restricted to a finely tuned corner of the scanned region. It is widespread in both parameter directions and persists over a broad portion of the structure formation epoch, indicating that the onset of the pathological regime is a generic issue for this parametrization rather than an isolated accident. Second, since $f_{\rm MG}$ peaks at late times, the bound is most restrictive exactly where voids are expected to be deepest, after having had the longest time to evolve, and where observations are most informative. In this sense, the model fails in the most critical redshift range for void studies, as it rules out depths that are typical of low redshift voids.

We add a brief clarification on the physical meaning of the bounds in eqs.~\eqref{Eq:viability_criterion_voids_MG} and~\eqref{eq:minimum_allowed_void_depth}. From a physical perspective, if one filters the density field around a void center on very large scales, say $R\sim 100\,h^{-1}{\rm Mpc}$, the resulting $\delta_{\rm E}$ is not expected to approach values close to $-1$, simply because the average extends over a substantial volume~\cite{Verza:2019tvg,Bayer:2021iyb,Verza:2024rbm}. From this viewpoint, excluding models as soon as the maximum of $f_{\rm MG}$ exceeds unity may appear overly conservative. The key point is that decreasing the filtering scale and focusing on the inner regions, e.g. $R\sim 1\,h^{-1}{\rm Mpc}$, typically drives the enclosed contrast to more negative values, so $\delta_{\rm E}$ can naturally approach $-1$ even when the large scale average is modest. This is precisely why we discard models with $f_{\rm MG}>1$: even if a void is not particularly deep when characterized on large scales, physically plausible inner regions can still reach the depths required to trigger the pathological branch of the non-linear coupling, leading to an ill defined fifth force.

We now reexamine the pathological regime from a complementary perspective. The void-informed criterion in eq.~\eqref{Eq:viability_criterion_voids_MG} is a ``static'' requirement in the sense that it does not follow the evolution of a specific void solution. Instead, it monitors the time dependence of the background function $f_{\rm MG}(z)$ that controls the onset of the imaginary branch through the combination $1+f_{\rm MG}\,\delta_{\rm E}$. The logic is field based. At each redshift, the matter field can realize a broad range of local configurations, and we require that no physically admissible value of $\delta_{\rm E}$ can trigger an imaginary force anywhere. This motivates imposing eq.~\eqref{Eq:viability_criterion_voids_MG} as a conservative filter that removes any model for which the pathological branch can be reached by some allowed density configuration at some redshift. In what follows, we discuss a dynamical version of the same pathology, formulated directly in terms of when the hydrodynamical description ceases to be well defined along the evolution of an isolated void.

In GR, the hydrodynamical model breaks down at shell-crossing, i.e. $\delta_{\rm E}=\delta_{\rm E,sc}$. In MG models that admit an imaginary branch of the fifth force, the hydrodynamical evolution can break down for two independent reasons:
(i) shell-crossing, which marks the limit of validity of the single stream description and prevents continuing the evolution beyond $\delta_{\rm E,sc}$;
(ii) pathological onset, where $\mu_{\rm NL}$ becomes imaginary and the evolution ceases to be well defined. This provides a relevant obstruction only if it is encountered before shell-crossing. If the evolution remains real up to $\delta_{\rm E,sc}$, then the hydrodynamical description breaks down at shell-crossing, and the imaginary branch is never realized. Thus, a consistent assessment of the breakdown of the hydrodynamical description must account for both effects.

To quantify this interplay, we introduce three diagnostic thresholds, all defined as functions of $z$ and extracted from the hydrodynamical evolution. The first quantity, $\delta_{\rm min,p}(z)$, where p denotes pathological, is the most negative density contrast that can be reached at redshift $z$ by solving the hydrodynamical equations while deliberately ignoring shell-crossing, under the requirement that the solution remains real and never enters the pathological regime. This procedure is well defined at the level of the differential equations, but it should not be interpreted as a physically consistent extension of the hydrodynamical model beyond its single stream regime. Depending on the model and the chosen redshift, the corresponding solution may or may not hit the imaginary branch. If it never does, then the deepest reachable value in this unphysical extension is simply $\delta_{\rm min,p}=-1$.
The second quantity, $\delta_{\rm min,h}(z)$, where h denotes hydrodynamical, translates the previous considerations into an effective limit on the hydrodynamical evolution by identifying the deepest value of $\delta_{\rm E}$ that can be reached at redshift $z$ once both breakdown mechanisms are taken into account.
Finally, the third quantity, $\delta_{\rm min,p_0}(z)$, isolates the specific evolution associated with the present day pathological threshold. We first compute $\delta_{\rm min,p}(z=0)$ and determine the corresponding critical IC $\delta_{\rm v,in,min,p_0}$ defined as
\begin{align}
        \delta_{\rm E}(z_{\rm in})=\delta_{\rm v,in,min,p_0}
        \quad\Rightarrow\quad
        \delta_{\rm E}(z=0)=\delta_{\rm min,p}(z=0)\,.
\end{align}
We then evolve the hydrodynamical equations using these ICs and record the resulting trajectory as a function of $z$. The value of this trajectory at each redshift defines $\delta_{\rm min,p_0}(z)$.

\begin{figure}
    \centering
    \includegraphics[width=1.0\linewidth]{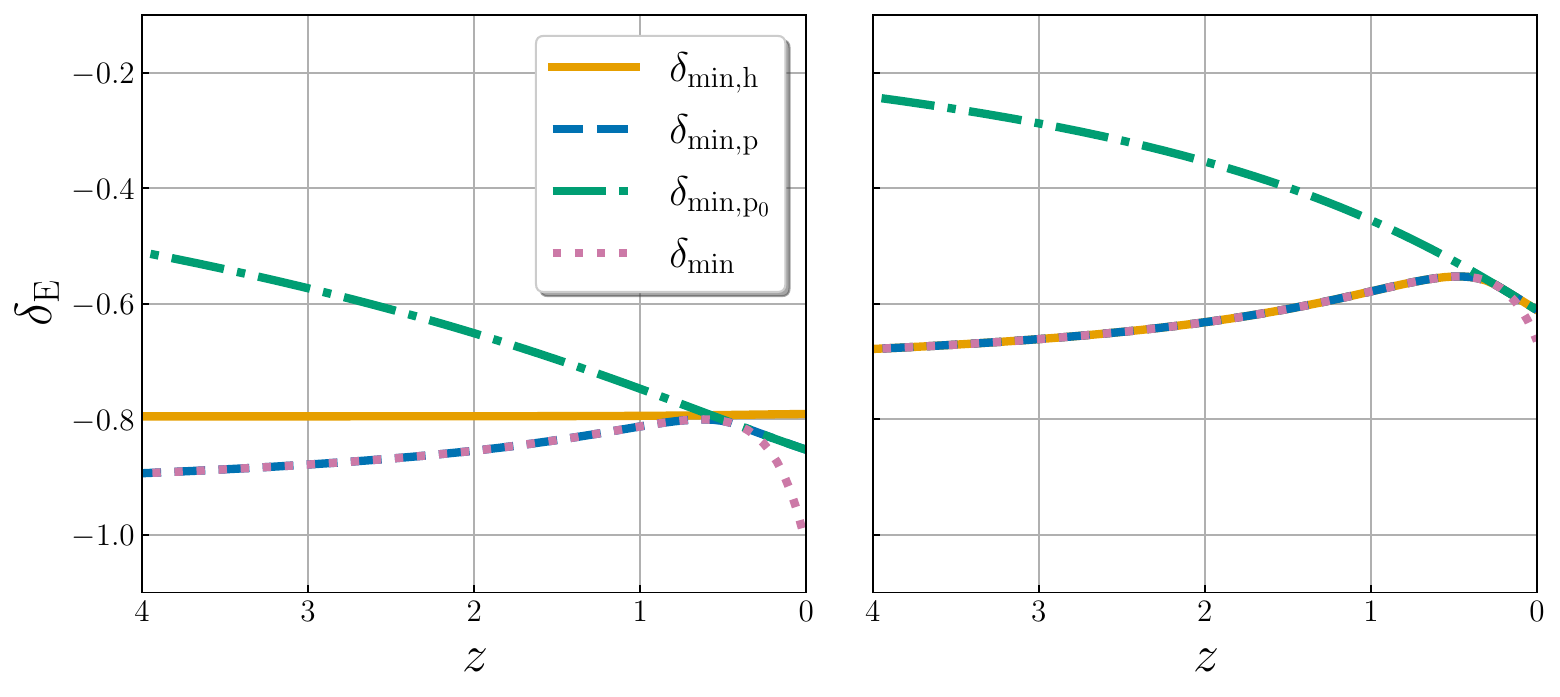}
    \caption{The evolution of $\delta_{\rm min}(z)$, $\delta_{\rm min,p}(z)$, $\delta_{\rm min,h}(z)$, and $\delta_{\rm min,p_0}(z)$ over $z\in[0,4]$ for two representative MG models. \textbf{Left:} $(\alpha_{\rm B_0},m)=(0.6,2.4)$. \textbf{Right:} $(\alpha_{\rm B_0},m)=(0.8,2.7)$.}
    \label{fig:max_depth_voids_comparison_two_models}
\end{figure}
In figure~\ref{fig:max_depth_voids_comparison_two_models} we plot $\delta_{\rm min}(z)$ together with the three diagnostics $\delta_{\rm min,p}(z)$, $\delta_{\rm min,h}(z)$, and $\delta_{\rm min,p_0}(z)$ over $z\in[0,4]$, for two representative parameter points: $(\alpha_{\rm B_0}=0.6,m=2.4)$ in the left panel, and $(\alpha_{\rm B_0}=0.8,m=2.7)$ in the right panel. Both models are excluded by the void-informed filter in eq.~\eqref{Eq:viability_criterion_voids_MG}.

A key feature common to both panels is the behavior of $\delta_{\rm min,p}(z)$ and its connection to $f_{\rm MG}(z)$. For the parametrization adopted in this work, $f_{\rm MG}(z)$ generally increases up to a maximum at $z_{\rm max}$ and decreases thereafter (see figure~\ref{fig:f_MG_delta_min}).
While $f_{\rm MG}(z)$ grows toward its maximum, $\delta_{\rm min}(z)$ increases accordingly, so the bound on the void depth becomes progressively more stringent, and the allowed window in $\delta_{\rm E}$ shrinks with time. As a result, $\delta_{\rm min,p}(z)=\delta_{\rm min}(z)$ for $z\geq z_{\rm max}$. This follows from monotonicity: since $\delta_{\rm min}(z)$ becomes less negative with time, while the void interior monotonically deepens, the deepest configuration allowed at redshift $z$ cannot have violated the bound at any earlier time. After the maximum, this identification no longer holds. In this regime, typically $\delta_{\rm min,p}(z)>\delta_{\rm min}(z)$, showing that the hydrodynamical evolution does not fail at redshift $z$. Instead, the relevant obstruction is set near $z_{\rm max}$, where the bound is tightest: any IC that would make the solution deeper than the critical trajectory at later times would have already crossed into the imaginary branch around $z_{\rm max}$. Hence, for $z\leq z_{\rm max}$, $\delta_{\rm min,p}(z)$ is simply the forward evolution of the solution tuned to sit at the boundary at $z_{\rm max}$, which also explains why $\delta_{\rm min,p_0}(z)$ coincides with $\delta_{\rm min,p}(z)$ in both panels. We include $\delta_{\rm min,p_0}(z)$ precisely to make this interpretation explicit.

Now, we can use the relative ordering of $\delta_{\rm min,h}(z)$ and $\delta_{\rm min,p}(z)$ to diagnose which mechanism actually halts the hydrodynamical evolution in a given model. The two panels then illustrate two limiting ways in which the hydrodynamical description can break down. In the left panel, one finds $\delta_{\rm min,h}(z)>\delta_{\rm min,p}(z)$ in the range shown, which implies that the hydrodynamical evolution is terminated by shell-crossing before the solution reaches the pathological branch of the fifth force. In this case, the imaginary branch is not encountered along the physically accessible hydrodynamical evolution, even though the underlying model is excluded by the void-informed criterion. However, in the right panel, $\delta_{\rm min,h}(z)=\delta_{\rm min,p}(z)$ shows that the hydrodynamical solution fails because it enters the pathological regime. These two examples bracket the spectrum of possible behaviors, and intermediate cases are expected within the broader parameter space.

After this detailed analysis, it is useful to highlight two take home messages. First, the existence of a regular hydrodynamical evolution down to shell-crossing over the redshift range relevant for LSS formation does not guaranty that the MG model is free of pathologies. As the left panel illustrates, one can have $\delta_{\rm min,p}(z)<\delta_{\rm min,h}(z)$ and still violate eq.~\eqref{Eq:viability_criterion_voids_MG}, because physically plausible inner regions can reach contrasts more negative than the shell-crossing threshold inferred from the single stream model.
Second, requiring that the hydrodynamical solution exists all the way down to shell-crossing over the probed redshift range does not guaranty viability and is systematically less restrictive than the void-informed criterion. Thus, it would, for instance, retain the left panel model in figure~\ref{fig:max_depth_voids_comparison_two_models}. This point is crucial for Void Size Function (VSF) models: although the hydrodynamical evolution is regular up to shell-crossing and would seemingly allow sensible VSF predictions, these predictions are not physically meaningful unless the general bound in eq.~\eqref{Eq:viability_criterion_voids_MG} is enforced, because the underlying MG model is in fact excluded and can admit realistic configurations that trigger an imaginary fifth force.

\section{Void evolution}
\label{sec:void_evolution}

We apply the hydrodynamical formalism to study the evolution of voids in MG, using as a concrete example the model introduced in section~\ref{sec:a_concrete_MG_model} and adopting the standard $w_0w_a$CDM scenario as a reference. Before presenting our results, let us briefly emphasize the main new ingredients of the analysis. Previous work developed the hydrodynamical description of spherical void evolution in GR and dynamical DE cosmologies~\cite{Moretti:2025gbp}, while related studies have investigated void-informed theoretical bounds in Galileon gravity~\cite{Moretti:2026axy}, spherical collapse and halo abundance in shift-symmetric Galileon theory~\cite{Albuquerque:2024hwv}, and the abundance of cosmic voids in the EFT of DE~\cite{Takadera:2025ehm}. The novelty of the present work is to formulate and apply a unified hydrodynamical treatment of non-linear void evolution in MG. In particular, we derive the modified void evolution equation in terms of the non-linear effective gravitational coupling, quantify the hierarchy between the linear and non-linear gravitational couplings in underdense regions, connect the pathological branch to the void-informed viability condition, and compute both the Lagrangian--Eulerian mapping and the shell-crossing thresholds within the same framework. These results provide the first application of the hydrodynamical void formalism to MG dynamics.

The results are organized as follows:
\begin{itemize}
    \item \textit{Impact of MG on single void evolution.}  
    In section~\ref{sec:impact_of_MG_on_single_void_evolution}, we study how the MG parameters $(\alpha_{\rm B_0},m)$ affect the evolution of an isolated void. We analyze $\mu_{\rm L}$ and $\mu_{\rm NL}$ and their dependence on these parameters. For the first time in this context, we derive the coupling hierarchy that controls screening in voids and its dependence on void depth. We then connect these ingredients to the corresponding evolution of $\delta_{\rm E}(z)$.
    
    \item \textit{Single void evolution with a pathological behavior.}  
    In section~\ref{sec:single_voids_evolution_with_a_pathological_behavior}, we implement the regularization used in $N$-body simulations~\cite{Barreira:2013eea} to handle the imaginary-force regime. For this test, we necessarily consider parameter choices that violate the viability bound in eq.~\eqref{Eq:viability_criterion_voids_MG}, and we use them to follow the transition from the physical branch to the regularized regime by monitoring both $\mu_{\rm NL}$ and $\delta_{\rm E}$.
    
    \item \textit{The map from Lagrangian to Eulerian space.}  
    In section~\ref{sec:the_map_from_Lagrangian_to_Eulerian_space}, we compute, for the first time, the mapping $\delta_{\rm v}(z,\delta_{\rm E})$ in MG and compare it to the corresponding $w_0w_a$CDM and EdS references.
    
    \item \textit{Shell-crossing.}  
    In section~\ref{sec:shell_crossing}, we apply the hydrodynamical shell-crossing condition to determine the non-linear and linearly extrapolated thresholds, $\delta_{\rm E,sc}(z)$ and $\delta_{\rm v,sc}(z)$, in MG. We then compare their redshift evolution across MG parameter choices and against the $w_0w_a$CDM and EdS reference cases.
\end{itemize}

All parameter choices considered in this section are well behaved, in the sense that they satisfy the viability bound in eq.~\eqref{Eq:viability_criterion_voids_MG}. The only exception is the illustrative test in the pathological regime, where we intentionally adopt excluded parameter choices to reproduce the $N$-body regularization prescription without drawing physical conclusions from those models. Moreover, as discussed in~\cite{Moretti:2025gbp}, earlier analyses often focused on void trajectories evolved down to shell-crossing. Here, we instead do not restrict ourselves to that limit, since void statistics can be defined for thresholds shallower than shell-crossing~\cite{Verza:2024rbm}, and the relevant information is not confined to configurations that reach it. Finally, throughout this section, we focus on spherical voids with an inverse top-hat density profile. In this configuration, the local and mean density contrasts coincide inside the void, so one can equivalently describe the dynamics in terms of $(\delta_{\rm E},\delta_{\rm L})$ or $(\Delta_{\rm E},\Delta_{\rm L})$. For consistency with the hydrodynamical formalism adopted in~\cite{Moretti:2025gbp} and in this work, we use $\delta_{\rm E}$ and $\delta_{\rm L}$ throughout.

\subsection{Impact of MG on single void evolution}
\label{sec:impact_of_MG_on_single_void_evolution}
In this section, we quantify the impact of MG on the evolution of an isolated, spherically symmetric void in a homogeneous and isotropic background, adopting an inverse top-hat as the initial density profile, i.e. eq.~\eqref{Eq:initial_top_hat_delta}. To assess the impact of MG, we evolve eqs.~\eqref{eq:delta_nl_MG_rewrite} and~\eqref{eq:delta_lin_MG_rewrite} for different values of $\alpha_{\rm B_0}$ and $m$, adopting identical ICs set deep in matter domination (see section~\ref{Eq:setup_for_isolated_void_evolution}), where MG and DE effects are negligible and all models approach the EdS limit. We choose these ICs such that, in the $w_0w_a$CDM case, the void reaches $\delta_{\rm E}=-0.5$ at $z=0$. This target depth is representative of the density contrasts of typical voids identified in $N$-body simulations and galaxy-survey catalogs~\cite{Verza:2019tvg,Contarini:2019qwf,Euclid:2022qtk,Massara:2022lng,Verza:2022qsh}. Keeping the ICs fixed while varying $\alpha_{\rm B_0}$ and $m$ ensures that any late-time departure from the $w_0w_a$CDM trajectory  traces the MG dynamics.

In this setup, the void evolution is characterized by three fundamental quantities: $\delta_{\rm E}$, $\mu_{\rm L}$, and $\mu_{\rm NL}$. To keep the discussion organized in a sequence of logically connected steps, we first focus on the evolution of the gravitational couplings, $\mu_{\rm L}$ and $\mu_{\rm NL}$, and their dependence on the MG parameters. We then build on this to discuss, for the first time, the behavior of the screening mechanism in voids. Then, we turn to the corresponding impact on the void solutions. This ordering is essential because studying the deviations in $\delta_{\rm E}$ and $\delta_{\rm L}$ requires understanding, first, the modifications in $\mu_{\rm L}$ and $\mu_{\rm NL}$, as well as how screening operates in voids.

\begin{figure}[t]
    \centering
    \includegraphics[width=1.0\linewidth]{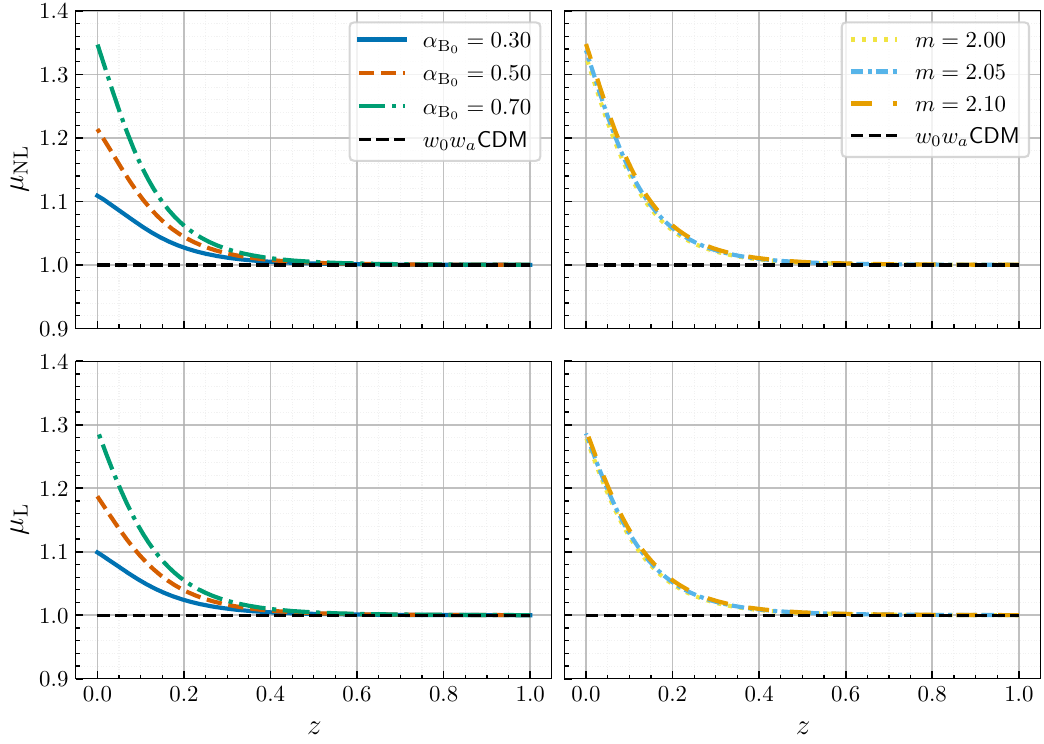}
    \caption{Non-linear $\mu_{\rm NL}$ (\textbf{top row}) and linear gravitational coupling $\mu_{\rm L}$ (\textbf{bottom row}) shown over $z\in[0,1]$. \textbf{Left column:} varying $\alpha_{\rm B_0}\in\{0.3,0.5,0.7\}$ at fixed $m=2.1$. \textbf{Right column:} varying $m\in\{2,2.05,2.10\}$ at fixed $\alpha_{\rm B_0}=0.7$. In all cases, the background parameters, including those of the reference $w_0w_a$CDM model, are fixed to the baseline cosmology specified in eq.~\eqref{eq:baseline_background}.}
    \label{fig:single_solution_couplings_varying_alpha_m}
\end{figure}

We present the results for the gravitational couplings in figure~\ref{fig:single_solution_couplings_varying_alpha_m}. The top row shows $\mu_{\rm NL}$, while the bottom row shows  $\mu_{\rm L}$. All quantities are plotted over $z\in[0,1]$, since in our parametrization, MG effects occur at late times. The left column varies $\alpha_{\rm B_0}\in\{0.3,0.5,0.7\}$ at fixed $m=2.1$, whereas the right column varies $m\in\{2,2.05,2.10\}$ at fixed $\alpha_{\rm B_0}=0.7$. In both columns, all remaining background parameters are kept fixed to the baseline cosmology in eq.~\eqref{eq:baseline_background}.

We note that both $\mu_{\rm NL}$ and $\mu_{\rm L}$ are strictly larger than one and tend to unity at early times. Increasing $\alpha_{\rm B_0}$ systematically amplifies the departures from GR ($\mu_{\rm NL}^{\rm GR}=\mu_{\rm L}^{\rm GR}=1$), while $\alpha_{\rm B_0}\to 0$ smoothly restores the GR limit. Similarly, increasing $m$ enhances the late-time modifications in both $\mu_{\rm NL}$ and $\mu_{\rm L}$, although the sensitivity to $\alpha_{\rm B_0}$ is noticeably stronger than that to $m$ across the ranges explored here. The departures from unity reach the ${\cal O}(10$--$35\%)$ level. This highlights why voids provide clean laboratories for MG: the effective gravitational coupling is substantially modified, and the internal dynamics, including halo and galaxy formation and peculiar motions within void regions, should reflect these deviations. Unlike overdense environments, where virialization complicates interpretation, voids keep expanding and probe progressively larger low-density volumes, providing favorable conditions for testing MG and assessing its impact on dynamics inside voids.

We now turn to the screening mechanism in voids and discuss its differences with respect to the spherical collapse case. Screening is captured by the hierarchy between $\mu_{\rm NL}$ and $\mu_{\rm L}$, that is, by how the non-linear gravitational response departs from the linear one.
We proceed in two complementary approaches. We first derive general analytic statements on the hierarchy of the gravitational couplings in void and halo configurations, which hold whenever $\mu_{\rm NL}$ takes the form given in eq.~\eqref{eq:non_linear_gravitational_potential}. We then apply these results to the specific void solutions discussed above and show how the strength of screening varies across the MG parameter space.

\begin{figure}
    \centering
    \includegraphics[width=1.0\linewidth]{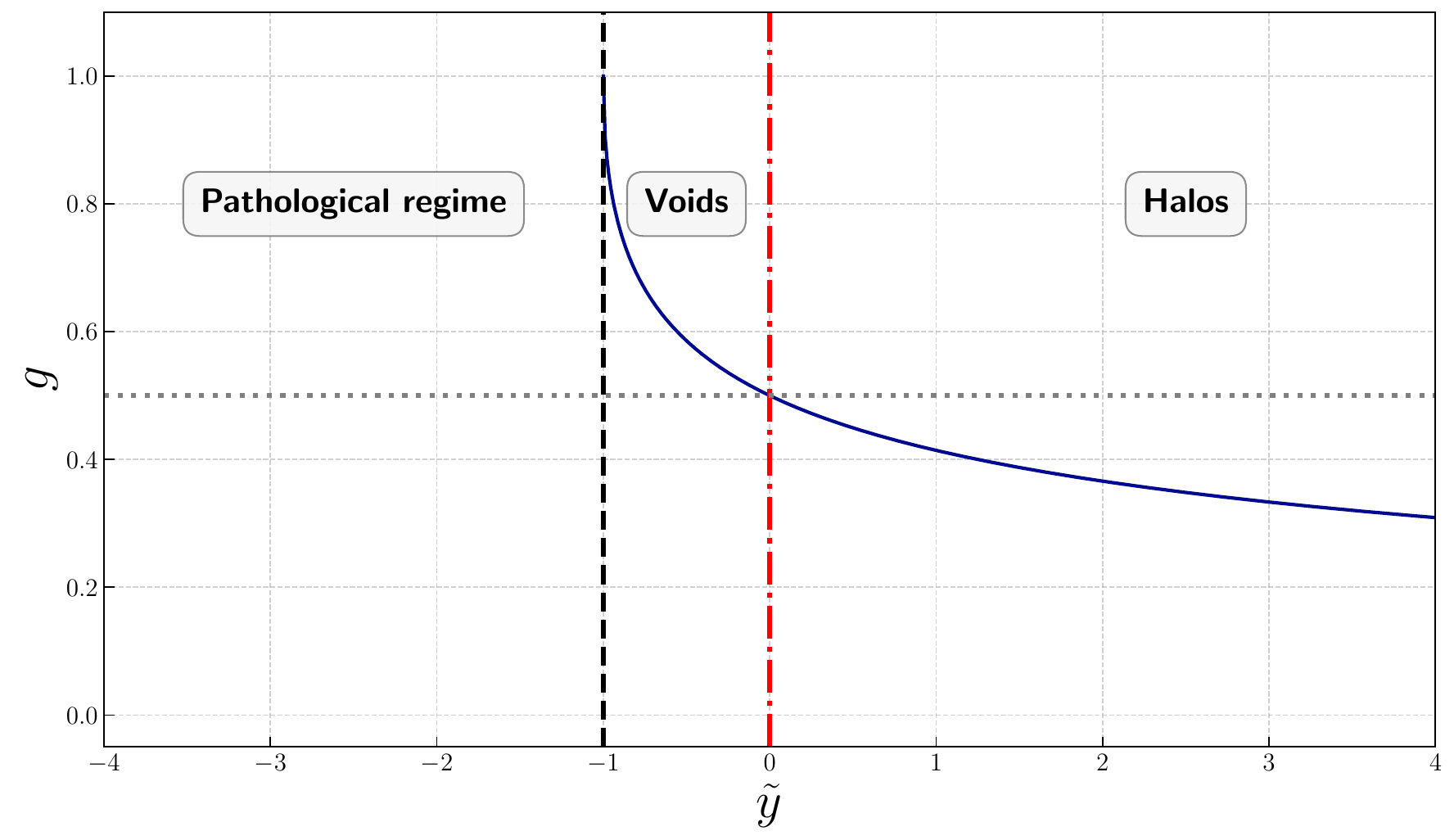}
    \caption{Auxiliary function $g(\tilde{y})$ defined in eq.~\eqref{eq:g_y}, shown over $\tilde{y}\in[-4,4]$. The intervals $\tilde{y}>0$ and $\tilde{y}\in[-1,0)$ correspond to halos and voids, respectively, while $\tilde{y}<-1$ marks the pathological regime where $g(\tilde{y})$ becomes imaginary. The point $\tilde{y}=0$ corresponds to the unperturbed background and serves as a limiting value, with no direct physical meaning as a separate configuration.}
    \label{fig:g_y}
\end{figure}
First of all, we introduce the auxiliary function
\begin{align}
    g(\tilde{y}) \equiv \frac{1}{2}\left(\frac{\mu_{\rm NL} - \upomega}{\mu_{\rm L}-\upomega}\right)\,=\,\frac{1}{\tilde{y}}\left[\sqrt{1+\tilde{y}}-1\right]\,, \qquad \tilde{y}=\left(\frac{R_{\rm V}}{R}\right)^3\,,\quad \upomega=\frac{M_{\rm pl}^2}{M^2}\,,
    \label{eq:g_y}
\end{align}
which is a natural choice to study the hierarchy between the MG potentials, given the definition of $\mu_{\rm NL}$ as a function of $\mu_{\rm L}$ in eq.~\eqref{eq:non_linear_gravitational_potential}. For the parametrization adopted in eq.~\eqref{eq:alphaB_benchmark}, $\upomega=1$. We show $g(\tilde{y})$ over $\tilde{y}\in[-4,4]$ in figure~\ref{fig:g_y}. We note that $g(\tilde{y})$ is monotonic and decreasing; it reaches its maximum value $g(-1)=1$, approaches $g(0)=0.5$ in the smooth limit $\tilde{y}\to 0$, and vanishes asymptotically as $\tilde{y}\to+\infty$. To translate these properties into relations between the gravitational couplings, we notice that $\tilde{y}>0$ corresponds to overdensities, that is, halos ($\delta_{\rm E}>0$ in eq.~\eqref{eq:RV_tophat_void_merged}), while voids correspond to $\tilde{y}<0$, and the physical non-pathological branch is $y\in[-1,0)$. The region $\tilde{y}<-1$ belongs to the pathological branch discussed in section~\ref{sec:viability}, where the square root in $g(\tilde{y})$ becomes imaginary, and the non-linear force ceases to be physically admissible. This regime is never reached by the models and parameter choices considered in the present analysis. It then follows that the linear and non-linear couplings satisfy the hierarchy
\begin{align}
    \textbf{Voids:}\qquad      &\tilde{y}\in[-1,0) \quad &\Longrightarrow\quad &\mu_{\rm L}<\mu_{\rm NL}\leq 2\,\mu_{\rm L}- \upomega\,,\\
    \textbf{Halos:}\qquad      &\tilde{y}>0        \quad &\Longrightarrow\quad &\mu_{\rm L}>\mu_{\rm NL}\,.
    \label{eq:gravitational_couplings_hierarchy}
\end{align}
The net outcome of this analysis is that screening in halos always suppresses the non-linear modifications relative to the linear ones (see, e.g.,~\cite{Albuquerque:2024hwv}, where the same MG model used here is investigated), whereas voids exist in an unscreened regime, and the non-linear coupling is always strictly larger than the linear one. As a result, MG effects are generally more pronounced in void environments than in halos. Finally, we notice that the upper bound on $\mu_{\rm NL}$ in the voids can be interpreted as  the requirement that the theory remains well behaved; i.e., one must stay on the non-pathological branch. This statement is meant in a qualitative sense since the onset of the imaginary branch cannot be mapped one to one onto a single condition expressed only in terms of $\mu_{\rm NL}$.

\begin{figure}[t]
    \centering
    \includegraphics[width=1.0\linewidth]{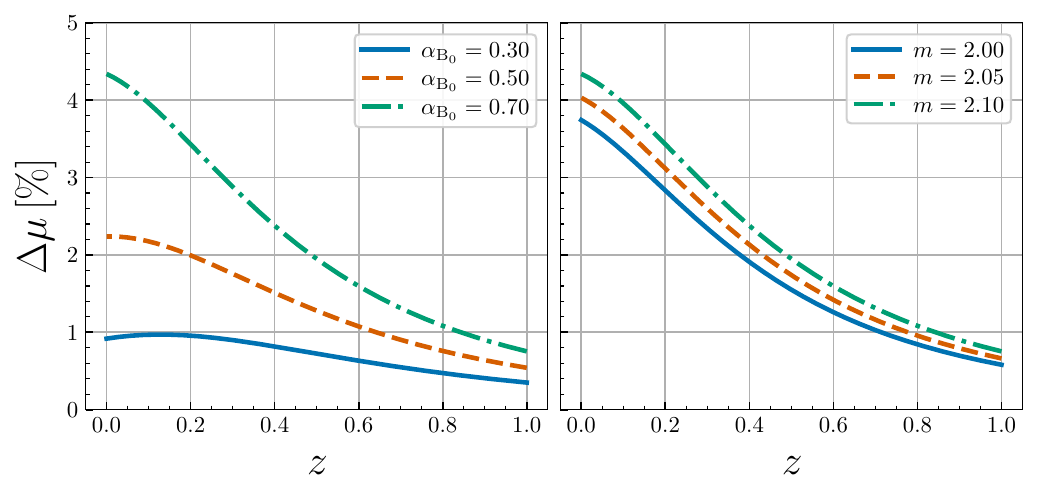}
    \caption{Percent relative difference of $\mu_{\rm NL}$ with respect to $\mu_{\rm L}$ as defined in eq.~\eqref{eq:percentdiffmuNLvsmuL} as function of redshift, when varying $\alpha_{\rm B_0}$ ({\bf left panel}) and $m$ ({\bf right panel}).}
    \label{fig:single_solution_perc_diff_mu}
\end{figure}
Then, to connect the analytic hierarchy in eq.~\eqref{eq:gravitational_couplings_hierarchy} to our void solutions and to quantify how screening varies across the MG parameter space, we show in figure~\ref{fig:single_solution_perc_diff_mu} the relative percentage difference between the non-linear and linear gravitational couplings for the model under consideration, defined as follows:
\begin{align}\label{eq:percentdiffmuNLvsmuL}
    \Delta\mu[\%] \,=\, \frac{\mu_{\rm NL}-\mu_{\rm L}}{\mu_{\rm L}}\,\times\,100 \,.
\end{align}
We find relative percentage differences at the $1$--$4\%$ level, indicating an amplified MG interaction in the non-linear gravitational coupling function. This shows that, in the models considered, cosmic voids are largely unscreened. This unscreened behavior becomes more pronounced as the MG parameters are increased: larger $\alpha_{\rm B_0}$ or $m$ enhance the MG modifications and drive a larger relative percentage difference between $\mu_{\rm NL}$ and $\mu_{\rm L}$.

\begin{figure}
    \centering
    \includegraphics[width=1.0\linewidth]{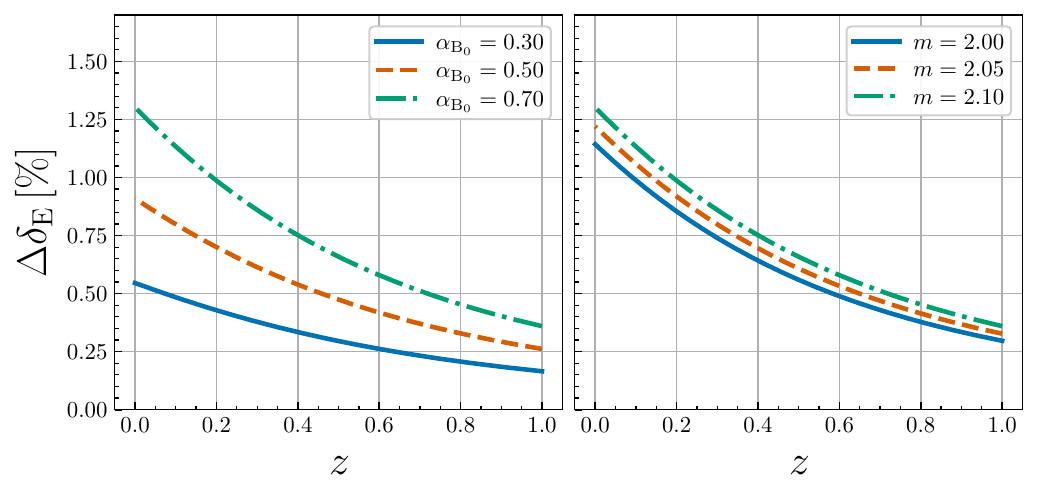}
    \caption{Percent relative difference between MG and $w_0w_a$CDM in the redshift evolution of $\delta_{\rm E}$, varying $\alpha_{\rm B_0}$ (left panel) and $m$ (right panel).}
    \label{fig:single_solution_relative_perc_difference}
\end{figure}
We are now ready to assess how MG impacts the density evolution inside voids, i.e. $\delta_{\rm E}$. In figure~\ref{fig:single_solution_relative_perc_difference}, we plot the relative percentage difference between the MG models under consideration and the $w_0w_a$CDM solutions, defined as
\begin{align}
    \Delta\delta_{\rm E}[\%] \,=\, \frac{\delta_{\rm E}-\delta_{\rm E}^{w_0w_a{\rm CDM}}}{\delta_{\rm E}^{w_0w_a{\rm CDM}}}\,\times\,100 \,.
\end{align}
The deviations remain at the percent level across the explored parameter space.
The sign of $\Delta\delta_{\rm E}[\%]$ is determined by $\mu_{\rm NL}$ in eq.~\eqref{eq:delta_nl_GR_rewrite}: enhancing the effective Newtonian coupling deepens the void with respect to the $w_0w_a$CDM case. The dependence on MG parameters is consistent with the behavior of the gravitational coupling: the larger $\alpha_{\rm B_0}$ and $m$ are, the larger the departure becomes.
The fact that $\Delta\delta_{\rm E}[\%]$ remains modest compared to the ${\cal O}(20$--$30\%)$ deviations in the gravitational couplings reflects the late-time nature of the MG corrections in our parametrization: the modification acts over a relatively short redshift interval and does not integrate into a large cumulative change in the density evolution. Nevertheless, percent-level shifts in void profiles are potentially within reach of ongoing and upcoming surveys such as \emph{Euclid}~\cite{Euclid:2021xmh,Euclid:2022qtk,Euclid:2022hdx,Euclid:2023eom}.
We note that in literature the dynamics is generally solved without non-linear and no-screening effects in the gravitational potential~\cite{Voivodic:2016kog,Falck:2017rvl,Clampitt:2012ub,Barreira:2015vra} leading to an underestimate of MG effects. In appendix~\ref{Sec:non_linear_void_dynamics_with_linear_gravitational_coupling}, we quantify this difference.

\begin{figure}
    \centering
    \includegraphics[width=1.0\linewidth]{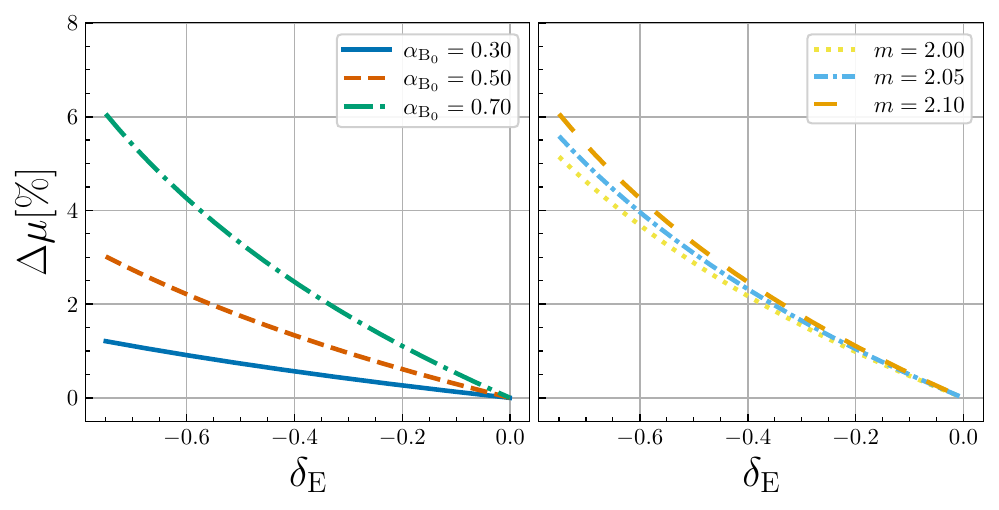}
    \caption{Percentage deviation of the non-linear gravitational coupling $\mu_{\rm NL}(z=0)$ with respect to the linear one, $\mu_{\rm L}(z=0)$, as a function of the final void depth $\delta_{\rm E}\in[-0.75,-0.001]$. \textit{Left panel:} variations of $\alpha_{\rm B_0}\in\{0.3,0.5,0.7\}$. \textit{Right panel:} variations of $m\in\{2.00,2.05,2.10\}$. In each case, all remaining cosmological and MG parameters are held fixed to the reference model.
}
    \label{fig:mu_nl_vs_delta_z_fixed}
\end{figure}

We conclude this section by analyzing how the non-linear gravitational couplings depend on the final void depth $\delta_{\rm E}$. In figure~\ref{fig:mu_nl_vs_delta_z_fixed}, we plot the percentage deviation of $\mu_{\rm NL}$ from the linear prediction at $z=0$ as a function of $\delta_{\rm E}\in[-0.75,-0.001]$. Specifically, we show
\begin{align}
    \Delta \mu[\%](z=0,\delta_{\rm E}) \,=\, \frac{\mu_{\rm NL}(z=0,\delta_{\rm E})-\mu_{\rm L}(z=0)}{\mu_{\rm L}(z=0)}\,\times\,100 \,,
\end{align}
where the left panel fixes $m=2.1$ and varies $\alpha_{\rm B_0}\in[0.3,0.5,0.7]$, while the right panel fixes $\alpha_{\rm B_0}=0.7$ and varies $m\in[2,2.05,2.1]$. The background parameters are fixed to the baseline model in eq.~\eqref{eq:baseline_background}.

For shallow underdensities, $\delta_{\rm E}\to 0^{-}$, the non-linear coupling smoothly approaches the linear prediction, $\mu_{\rm NL}\to\mu_{\rm L}$, so that the fractional deviation becomes negligible. As the void deepens, $\Delta\mu[\%]$ increases monotonically with $|\delta_{\rm E}|$, signaling a progressive departure of $\mu_{\rm NL}$ from $\mu_{\rm L}$ in the increasingly non-linear regime. This growth with $|\delta_{\rm E}|$ is present in all models and spans from the sub-percent level at mild depths to the percent level and, for the strongest MG choices, up to ${\cal O}(6\%)$ for the deepest voids. What changes across parameter choices is primarily the rate at which $\Delta\mu[\%]$ grows with $|\delta_{\rm E}|$: stronger MG leads to a steeper $\delta_{\rm E}$-dependence and thus a larger sensitivity of $\mu_{\rm NL}$ to void depth. 

\subsection{Single void evolution with a pathological behavior}
\label{sec:single_voids_evolution_with_a_pathological_behavior}

In section~\ref{sec:viability}, we argued that if, during the redshift range relevant for structure formation, a void configuration renders the fifth force imaginary, then the underlying MG model is physically inconsistent and is discarded. Nevertheless, some Galileon $N$-body simulations~\cite{Barreira:2013eea} still explore models that would be excluded by the viability criterion in eq.~\eqref{Eq:viability_criterion_voids_MG}, running them by adopting an \textit{ad hoc} prescription inside voids whenever the fifth force would otherwise become imaginary. Thus, the goal of this section is threefold: (i) to document how we implement the same \textit{ad hoc} prescription within our spherical hydrodynamical model, enabling a direct comparison with existing $N$-body results; (ii) to test whether the approach to the would-be pathological threshold leaves any precursor signatures in $\delta_{\rm E}$ or $\mu_{\rm NL}$, indicating that the evolution is nearing the imaginary force regime; and (iii) to characterize the transition itself, i.e. how the solution switches from the physical branch to the regularized one. Note that points (ii) and (iii) are fully compatible: the regularized evolution still allows us to probe the neighborhood of the threshold and to identify potential precursors that would precede the onset of the pathology in the unregularized dynamics.

To this end, we first summarize the specific regularization employed in these simulations and describe its implementation in our framework. The regularization (see~\cite{Barreira:2013eea}) is imposed directly on the ratio $R_{\rm V}/R$, since this quantity can trigger the onset of the pathological (imaginary force) regime: whenever the evolution would push it past the threshold, the defining relation is saturated to enforce a real-valued branch for the square root in eq.~\eqref{eq:non_linear_gravitational_potential}. Concretely, the prescription reads
\begin{align}
    \left(\frac{R_{\rm V}}{R}\right)^3 = \max\!\left[-1\,, f_{\rm MG}(z)\,\delta_{\rm E}(z)\right] \,.
    \label{eq:prescription_N_body_imaginary_force}
\end{align}
Whenever $f_{\rm MG}\,\delta_{\rm E}<-1$ (pathological regime), the right-hand side is saturated at $-1$, which forces the fifth force to remain real and implies $\mu_{\rm NL}=2\mu_{\rm L}-1$. This connects directly to the discussion in section~\ref{sec:impact_of_MG_on_single_void_evolution}: qualitatively, a violation of $\mu_{\rm NL} \leq 2\mu_{\rm L}-1$ is precisely what one expects as the evolution approaches the configuration in which $R_{\rm V}/R$ would trigger the imaginary branch.  In practice, we implemented eq.~\eqref{eq:prescription_N_body_imaginary_force} by replacing the original expression for $(R_{\rm V}/R)^3$ with the max-prescription above, so that our hydrodynamical evolution matches the force modeling adopted in the reference $N$-body simulation.

\begin{figure}
    \centering\includegraphics[width=1.0\linewidth]{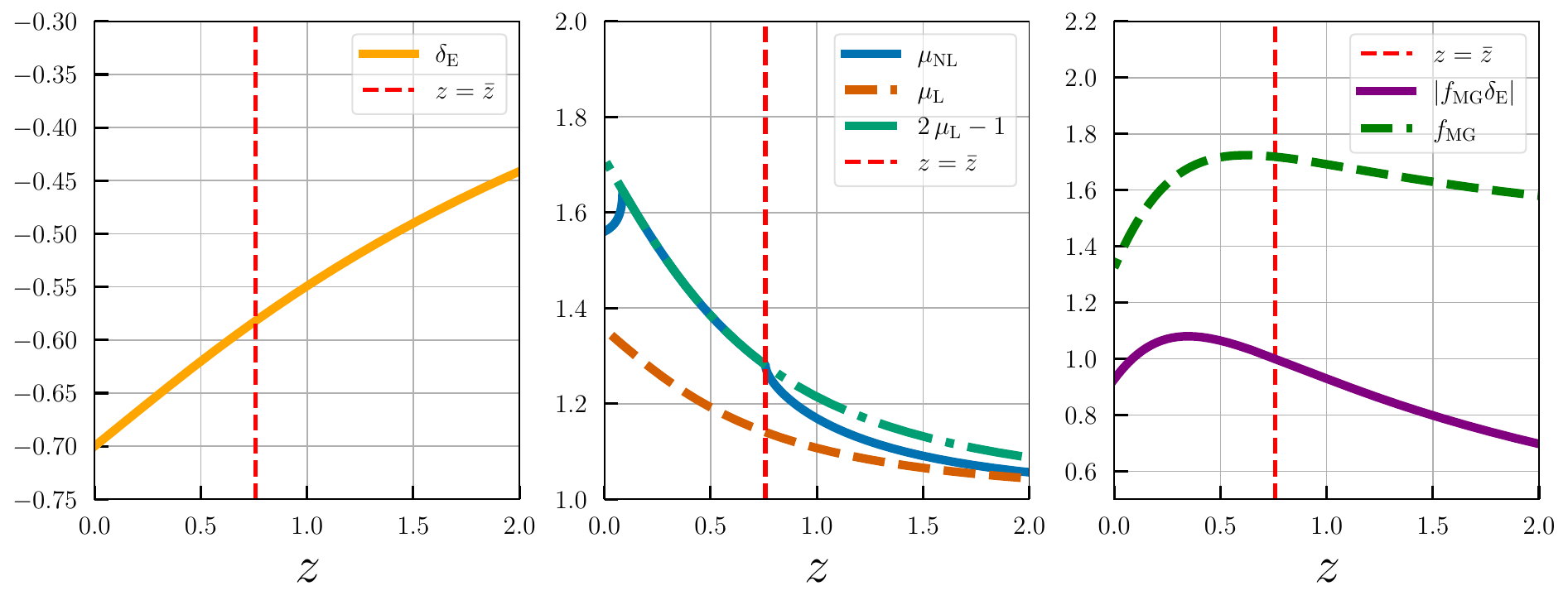}
    \caption{
    Isolated void evolution in the excluded (see figure~\ref{fig:max_funz_sqrt}) Galileon model $(w_0,w_a,\alpha_{\rm B_0},m)=(-0.97,-0.11,0.7,2.7)$, tuned to reach $\delta_{\rm E}(z=0)=-0.7$, with the $N$-body-inspired regularization prescription applied to $\mu_{\rm NL}$ to avoid the imaginary force regime. Left: $\delta_{\rm E}(z)$; the vertical dashed line marks $z=\bar{z}$, the first redshift where $f_{\rm MG}\,\delta_{\rm E}<-1$. Middle: $\mu_{\rm NL}$, $\mu_{\rm L}$, and $2\mu_{\rm L}-1$. Right: $f_{\rm MG}$ and $|f_{\rm MG}\delta_{\rm E}|$. All quantities are shown over the same redshift interval $z\in[0,2]$. 
}    \label{fig:single_imaginary_force_assumption}
\end{figure}
We now use this controlled setup to inspect the behavior of a void solution that would cross into the pathological regime in the absence of regularization. We consider the excluded model (see figure~\ref{fig:max_funz_sqrt}) with $(\alpha_{\rm B_0},m)=(0.7,2.7)$. We select a trajectory reaching \mbox{$\delta_{\rm E}(z=0)=-0.7$} and determine the corresponding ICs through a shooting procedure performed with the regularized force, ensuring that the evolution encounters the would-be pathological threshold at some $\bar{z}>0$. The results are shown in figure~\ref{fig:single_imaginary_force_assumption}. The left panel displays $\delta_{\rm E}(z)$ over $z\in[0,2]$; the vertical dashed red line marks $z=\bar{z}$, defined as the largest redshift at which $f_{\rm MG}\,\delta_{\rm E}<-1$ (i.e. the first entry into the pathological regime when evolving from high to low redshift). The same vertical line is also shown in the central and right panels for reference. The central panel shows $\mu_{\rm NL}$, $\mu_{\rm L}$, and $(2\mu_{\rm L}-1)$ over the same interval, while the right panel reports $f_{\rm MG}$ and $|f_{\rm MG}\,\delta_{\rm E}|$.

Two points emerge from the left panel. First, for $z>\bar{z}$, i.e. before the first pathological entry, $\delta_{\rm E}$ remains smooth and displays no sign of non-linear breakdown. This differs from the analysis of~\cite{Winther:2015pta}, where the fully dynamical, non-QSA evolution breaks down at the same stage where the QSA branch becomes ill defined, with the numerical solution becoming unstable and blowing up. Here, instead, the pathological regime is still encountered at the level of $\mu_{\rm NL}$, while the density evolution remains regular and shows no precursor instability before $\bar{z}$. Second, the solution remains continuous also for $z<\bar{z}$: the void keeps expanding and evacuating matter, as one expects for void evolution, with no discontinuities or other strange features in $\delta_{\rm E}$ induced by the prescription in eq.~\eqref{eq:prescription_N_body_imaginary_force}.

The behavior of the non-linear gravitational coupling is more subtle. The central panel makes explicit that the evolution satisfies $\mu_{\rm L} < \mu_{\rm NL} < 2\mu_{\rm L}-1 \,,$ until the $\bar{z}$ is reached, after which $\mu_{\rm NL}$ is driven onto the saturated branch $\mu_{\rm NL}=2\mu_{\rm L}-1$ whenever $f_{\rm MG}\,\delta_{\rm E}<-1$. Importantly, $\mu_{\rm NL}$ remains continuous across the transitions, but the switching between the two branches generates cusp-like points: the time derivative of $\mu_{\rm NL}$ is not the same when approached from the left or from the right. In the present example, two such cusps are visible, indicating two distinct switches between the physical branch and the saturated branch enforced by the prescription, namely the regime where $\mu_{\rm NL}$ is set to $2\mu_{\rm L}-1$, corresponding to the would-be pathological regime. Consistently, the right panel shows that at both transition points $|f_{\rm MG}\,\delta_{\rm E}|=1$. 

Finally, the right panel clarifies another important point: entering the pathological regime once, does not necessarily imply remaining there permanently.
Since the condition is controlled by $f_{\rm MG}(z)\,\delta_{\rm E}(z)$, a sufficiently strong late-time decrease in $f_{\rm MG}$ can bring the system back into the well-behaved regime, even while the void continues to expand. This also explains the non-monotonic behavior of $\mu_{\rm NL}$: even though $\mu_{\rm NL}>\mu_{\rm L}$ for this Galileon parametrization, there is no reason for $\mu_{\rm NL}$ to increase monotonically with time, since its detailed evolution is governed by $f_{\rm MG}(z)$.

\subsection{The map from Lagrangian to Eulerian space}
\label{sec:the_map_from_Lagrangian_to_Eulerian_space}

In this section, we assess the impact of MG models on the mapping between Lagrangian and Eulerian spaces.

We recall the physical meaning of the map. Given a void characterized by a non-linear matter density contrast $\delta_{\rm E}$ at redshift $z$, the mapping returns the corresponding linearly extrapolated value $\delta_{\rm v}(z,\delta_{\rm E})$, namely the density contrast that the same configuration would have at the same redshift if it had evolved according to linear theory. In practice, $z$ and $\delta_{\rm E}$ can be specified freely, provided that $\delta_{\rm E}(z)\geq\delta_{\rm E,sc}(z)$, since configurations deeper than shell-crossing at that redshift are theoretically inconsistent, as discussed in~\cite{Moretti:2025gbp}. For the numerical construction of $\delta_{\rm v}(z,\delta_{\rm E})$, as well as for its use in modeling the VSF, we refer the reader to~\cite{Moretti:2025gbp}.

\begin{figure}
    \centering
    \includegraphics[width=1.0\linewidth]{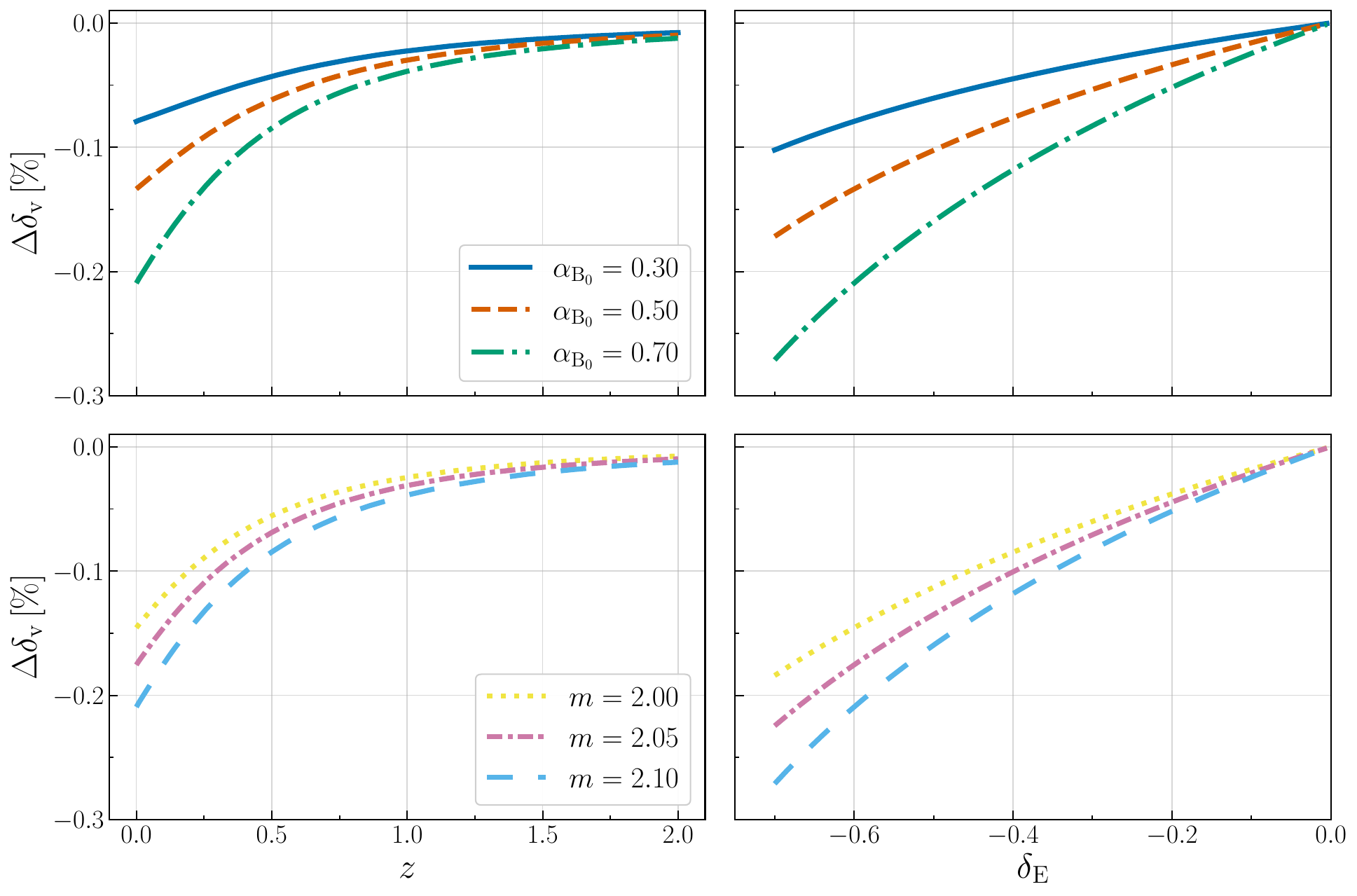}
    \caption{We present the relative percentage difference on the mapping, namely $\Delta\delta_{\rm v}[\%]$, between the $w_0w_a$CDM model and the corresponding MG realization. Left column: $\Delta\delta_{\rm v}[\%]$ as a function of $z\in[0,2]$ at fixed $\delta_{\rm E}=-0.6$. Right column: $\Delta\delta_{\rm v}[\%]$ as a function of $\delta_{\rm E}\in[-0.7,-0.001]$ at fixed $z=0$. Top row: varying $\alpha_{\rm B_0}\in\{0.3,0.5,0.7\}$ at fixed $m=2.1$. Bottom row: varying $m\in\{2.00,2.05,2.10\}$ at fixed $\alpha_{\rm B_0}=0.7$.}
    \label{fig:map_z_d}
\end{figure}
We now quantify the sensitivity of the map to the MG parameters using figure~\ref{fig:map_z_d}. In each panel, we show the percentage deviation with respect to the $w_0w_a$CDM reference mapping, i.e.
\begin{align}
    \Delta\delta_{\rm v}[\%] \equiv
    \frac{\delta_{\rm v}^{\rm model}(z,\delta_{\rm E})-\delta_{\rm v}^{w_0w_a{\rm CDM}}(z,\delta_{\rm E})}
    {\delta_{\rm v}^{w_0w_a{\rm CDM}}(z,\delta_{\rm E})}\times100\,,
\end{align}
where $\delta_{\rm v}^{w_0w_a{\rm CDM}}(z,\delta_{\rm E})$ denotes the mapping computed in a $w_0w_a$CDM model with the same background cosmological parameters adopted in eq.~\eqref{eq:baseline_background}, while $\delta_{\rm v}^{\rm model}(z,\delta_{\rm E})$ is the corresponding MG prediction.

The left column fixes $\delta_{\rm E}=-0.6$ and shows the redshift dependence over $z\in[0,2]$. The right column fixes $z=0$ and shows the dependence on void depth over $\delta_{\rm E}\in[-0.7,-0.001]$. We have explicitly verified that, for all models considered and over the full redshift range shown, these choices remain above the shell-crossing threshold $\delta_{\rm E,sc}(z)$. This plotting choice, in which we fix one argument of the map while varying the other, cleanly isolates the response of $\delta_{\rm v}(z,\delta_{\rm E})$ to changes in $z$ versus $\delta_{\rm E}$ and improves the readability of MG effects. In the left column, we vary $\alpha_{\rm B_0}\in\{0.3,0.5,0.7\}$ at fixed $m=2.1$, while in the right column we vary $m\in\{2.00,2.05,2.10\}$ at fixed $\alpha_{\rm B_0}=0.7$.

Across all panels, the deviation remains subpercent, both when varying $z$ at fixed $\delta_{\rm E}$ and when varying $\delta_{\rm E}$ at fixed $z$. Moreover, at high redshift $\Delta\delta_{\rm v}[\%]\to0$, consistent with the fact that MG effects switch off and the dynamics approaches the $w_0w_a$CDM limit. A common feature of all curves is that $\Delta\delta_{\rm v}[\%]$ is negative, implying
\begin{align}
    \delta_{\rm v}^{\rm MG}(z,\delta_{\rm E})>\delta_{\rm v}^{\rm GR}(z,\delta_{\rm E}).
\end{align}
The sign can be understood by the same logic discussed in~\cite{Moretti:2025gbp}. In the left column, we compare voids with the same $\delta_{\rm E}$ at the same redshift across different models. Two competing ingredients enter the mapping. First, since gravity is effectively stronger in MG $(\mu_{\rm NL}>1)$, the non-linear evolution reaches a given final underdensity more efficiently; thus, matching a fixed $\delta_{\rm E}$ requires larger ($\delta_{\rm E}<0$) ICs in MG than in GR. Second, the linear growth is also enhanced in MG, which tends to increase the linear extrapolation of a given initial perturbation and therefore acts in the opposite direction. The results indicate that the enhancement of linear growth does not compensate for the shift in the required ICs. Consistently, the deviation increases toward low redshift, where the MG prescription adopted here produces its largest modifications, and it vanishes at high redshift as the dynamics approach the GR limit.

In the right column, the idea is similar. Here we compare $\delta_{\rm v}(z=0,\delta_{\rm E})$ across models while varying $\delta_{\rm E}$. Since the comparison is performed at $z=0$, the linear growth history is fixed within each model and does not generate any additional $\delta_{\rm E}$ dependence in the relative deviation. What changes with void depth is the non-linear matching: we observe that the shift in the ICs required to reproduce a given final $\delta_{\rm E}$ becomes progressively larger for deeper voids. This trend is consistent with the fact that the relative difference between the non-linear and linear gravitational couplings increases in magnitude as $\delta_{\rm E}$ decreases, as shown in figure~\ref{fig:mu_nl_vs_delta_z_fixed}. Therefore, MG effects are amplified in deeper voids, and the larger non-linear shifts directly propagate into a stronger deviation of the mapping, explaining why $|\Delta\delta_{\rm v}[\%]|$ increases as $\delta_{\rm E}$ decreases.

Finally, the MG parameter dependence of the curves is consistent with this picture. As either $\alpha_{\rm B_0}$ or $m$ is increased, the MG curves move farther away from zero.  This behavior supports the interpretation that stronger effective gravity enhances the imbalance between the non-linear shift in the required ICs and the corresponding change in linear growth, thereby increasing the net impact on $\delta_{\rm v}(z,\delta_{\rm E})$.

Although they remain at the subpercent level, these deviations may still propagate into non-negligible changes in the theoretical VSF. This is analogous to spherical collapse, where small variations in the collapse threshold induce percent-level shifts in the HMF (see, e.g.,~\cite{Pace:2010sn}). A comparable sensitivity is therefore expected for void statistics. We therefore conclude that even small but coherent modifications of the mapping can matter when confronting theory with data in high-precision void analyses.

\begin{figure}
    \centering
    \includegraphics[width=1.0\linewidth]{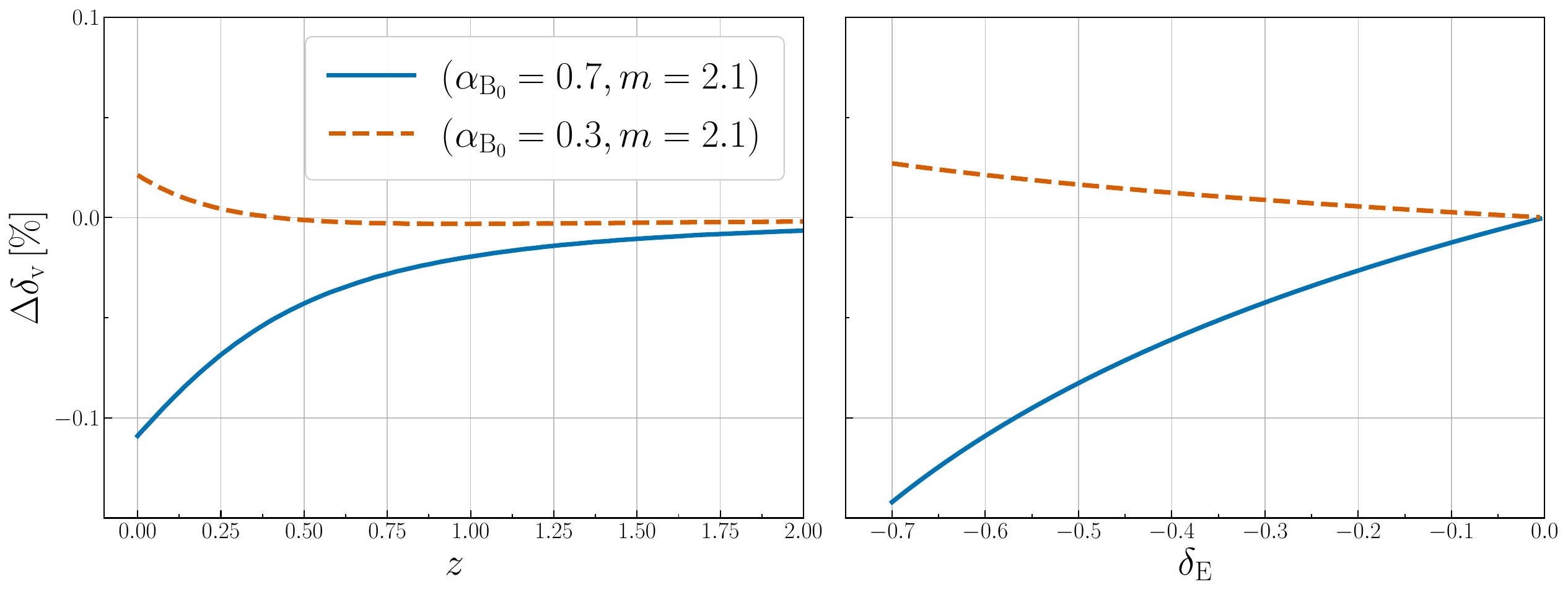}
    \caption{Relative percentage difference on the mapping, $\Delta\delta_{\rm v}[\%]$, between two MG models and the EdS reference. Left panel: varying $z$ at fixed $\delta_{\rm E}=-0.6$. Right panel: varying $\delta_{\rm E}$ at fixed $z=0$. The MG parameter points are $(\alpha_{\rm B_0},m)=(0.7,2.1)$ and $(\alpha_{\rm B_0},m)=(0.3,2.1)$.}
    \label{fig:comparison_MG_EdS}
\end{figure}
We conclude this section with figure~\ref{fig:comparison_MG_EdS}, which follows the same layout as figure~\ref{fig:map_z_d} but compares two MG models, $(\alpha_{\rm B_0},m)=(0.7,2.1)$ and $(\alpha_{\rm B_0},m)=(0.3,2.1)$, to EdS rather than to the $w_0w_a$CDM reference.
The results can be interpreted in terms of the same two competing mechanisms discussed above. For a fixed final $\delta_{\rm E}$, the ICs are less negative in EdS than in MG because structure formation in EdS is more efficient than in the MG realizations once the full expansion history is accounted for. The second ingredient is the linear growth, whose impact is always opposite to the non-linear shift in the required ICs. In particular, for $(\alpha_{\rm B_0},m)=(0.7,2.1)$ the change in linear growth overcompensates the initial shift, leading to $\delta_{\rm v}^{\rm MG}(z,\delta_{\rm E})>\delta_{\rm v}^{\rm EdS}(z,\delta_{\rm E})$. For $(\alpha_{\rm B_0},m)=(0.3,2.1)$ this behavior occurs only at high redshift in the left panel, while in all other cases the inequality is reversed. This qualitative difference arises because the MG models considered here combine two effects, namely a GR-like background with DE and a modified gravitational coupling at the perturbation level, so the relative balance between non-linear matching and linear growth is both parameter and redshift dependent.

\subsection{Shell-crossing}
\label{sec:shell_crossing}

In this section, we present the final results of our analysis, namely the non-linear and linear density contrasts at shell-crossing. All results are obtained within the hydrodynamical formalism by employing the shell-crossing condition derived in~\cite{Moretti:2025gbp}. For methodological details and implementation aspects, we refer the reader to that work. During the preparation of this manuscript,~\cite{Takadera:2025ehm} presented an independent analysis of the shell-crossing threshold in the same MG model considered here, adopting the $R$-based formulation and, consequently, the corresponding $R$-based shell-crossing condition introduced in~\cite{Moretti:2025gbp}. The two approaches are equivalent, and our thresholds are consistent with theirs.

\begin{figure}
    \centering
    \includegraphics[width=1.0\linewidth]{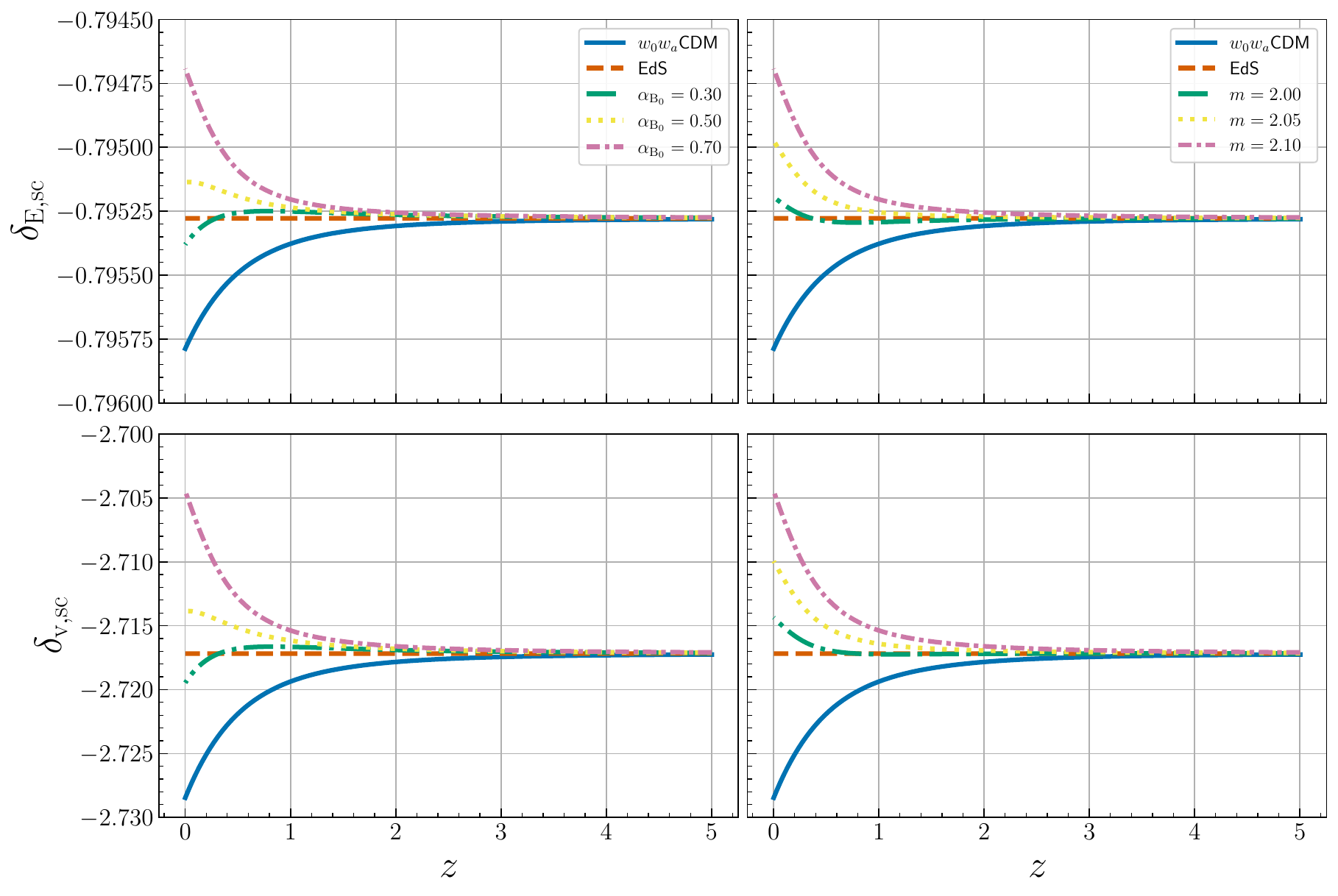}
    \caption{Redshift evolution of $\delta_{\rm E,sc}$ (top row) and $\delta_{\rm v,sc}(z)$ (bottom row) over $z\in[0,5]$, computed in the hydrodynamical formalism. In all panels, the background expansion is fixed to the baseline model in eq.~\eqref{eq:baseline_background}, and the MG parameter ranges match those adopted in the mapping analysis of the previous section. Left column: varying $\alpha_{\rm B_0}$ at fixed $m$. Right column: varying $m$ at fixed $\alpha_{\rm B_0}$. The dashed curve shows the results for the baseline $w_0w_a$CDM case.}
    \label{fig:shell_crossing}
\end{figure}
We present the results in figure~\ref{fig:shell_crossing}. The top row shows the non-linear density contrast at shell-crossing, $\delta_{\rm E,sc}$, while the bottom row shows the corresponding linearly extrapolated threshold, defined as $\delta_{\rm v,sc}\equiv\delta_{\rm v}(z,\delta_{\rm E,sc}(z))$. In both rows, we display the redshift evolution over $z\in[0,5]$, fixing the background expansion to the baseline one in eq.~\eqref{eq:baseline_background}. The MG parameters are varied over the same ranges adopted in the mapping analysis of the previous section. In the left column, we vary $\alpha_{\rm B_0}$ at fixed $m$, whereas in the right column, we vary $m$ at fixed $\alpha_{\rm B_0}$. For reference, we also show the $w_0w_a$CDM and EdS predictions.

The MG corrections to the shell-crossing thresholds are at the subpercent level with respect to the baseline $w_0w_a$CDM model. Although small, such shifts are theoretically relevant because they can propagate into percent level changes in the VSF, in close analogy with the role played by the spherical collapse thresholds in the HMF (see~\cite{Albuquerque:2024hwv}).

At high redshift, both $\delta_{\rm E,sc}$ and $\delta_{\rm v,sc}$ converge to the EdS values. This behavior follows from two simultaneous limits: the MG modifications vanish by construction, and the background approaches the matter dominated regime because the DE contribution becomes negligible. As a result, the dynamics reduces to the EdS.

From the curves, we see that MG systematically raises the shell-crossing threshold relative to GR, $\delta_{\rm E,sc}^{\rm MG}>\delta_{\rm E,sc}^{\rm GR}$. This behavior is expected because the enhanced non-linear coupling, $\mu_{\rm NL}>1$, speeds up the void evolution and amplifies the peculiar velocity field. Since the comparison is performed at fixed background cosmology, the larger void velocities in MG naturally make inner shells reach the same environment earlier. The dependence on $\alpha_{\rm B_0}$ and $m$ follows the same trend: larger values strengthen the MG modification and increase the separation from the GR curve.

A direct comparison to EdS is more subtle because two changes occur simultaneously: the background history differs, and MG strengthens gravity at the perturbation level. Since shell-crossing measures the catch-up of inner shells with the environment, the corresponding shift in the shell-crossing threshold is set by the interplay between the enhanced force and the different background expansion. Therefore, even if structure growth is faster in EdS, some MG realizations can still yield a shell-crossing threshold that is higher than the EdS one. In this case, a higher threshold does not imply that voids evolve more in MG than in EdS. It only indicates that, relative to their respective environments, inner shells in MG catch up with the background earlier than they do in EdS. Indeed, for fixed ICs, a void reaches shell-crossing first in EdS, then in MG, and finally in $w_0w_a$CDM.

The interpretation of the linear threshold, $\delta_{\rm v,sc}$, in MG relative to $w_0w_a$CDM or EdS follows the same picture discussed in the previous section.

\section{Conclusion}\label{sec:conclusion}

In this work, we developed a hydrodynamical description of the evolution of isolated spherical inverse top-hat cosmic voids in MG, in the QS regime, extending the GR/dynamical-DE framework introduced in our previous analysis~\cite{Moretti:2025gbp} to theories in which departures from GR can be encoded in effective non-linear modifications of the Poisson equation. Within this setup, the void dynamics is governed by the standard continuity and Euler equations, while MG enters through an effective gravitational coupling $\mu_{\rm NL}(a,R)$ in the non-linear evolution equation for the Eulerian density contrast $\delta_{\rm E}$ (and $\mu_{\rm L}$ at linear order). This provides a transparent and computationally efficient bridge between the gravitational coupling of a given MG model and late-time void observables.

Then, we specialized the formalism to an EFT of DE/MG description in the Galileon sub-class of models with luminal tensor speed and derived an explicit expression for $\mu_{\rm NL}$ in the QS regime, highlighting the role of derivative self-interactions through a Vainshtein scale. A key aspect of this class of models is the possible appearance of an unphysical ``imaginary'' fifth force in sufficiently underdense regions, signaled by the breakdown of the square-root branch in $\mu_{\rm NL}$. Building on the void-informed consistency criterion introduced in~\cite{Moretti:2026axy}, we used the background function controlling the onset of this pathology to identify the viable region for a benchmark model characterized by the $(\alpha_{\rm B_0},m)$ parameter space and to translate it into a redshift dependent bound on the minimum attainable void depth. We further introduced dynamical diagnostics that distinguish whether, for a given model, the hydrodynamical evolution is terminated first by shell-crossing (as in GR) or by the onset of the pathological branch; this discussion clarifies why requiring a regular evolution up to shell-crossing is not, by itself, a sufficient viability test. 

For parameter choices that satisfy the void-informed bound, we quantified the impact of MG on single void evolution starting from inverse top-hat ICs. In the explored EFT parametrization, both $\mu_{\rm L}$ and $\mu_{\rm NL}$ are enhanced at late times relative to GR, and we showed analytically that, in this theory class, voids always lie in an unscreened regime (while halos lie in a screened regime), implying $\mu_{\rm NL}>\mu_{\rm L}$ for the physical, non-pathological void branch. For the representative ranges considered, the resulting modification of the void density evolution remains at the percent level, despite $\mathcal{O}(10\text{--}30\%)$ changes in the effective couplings, reflecting the fact that the MG deviations become significant only at late times in this parametrization. We also characterized how the non-linear response of the gravitational coupling departs increasingly from the linear one as the final void depth increases.

Moreover, we provided the ingredients needed to connect ICs to late-time void observables beyond GR, including the mapping between Lagrangian and Eulerian descriptions and the determination of shell-crossing thresholds within the MG dynamics. An important qualitative outcome is that, in viable MG models with a stronger effective gravitational interaction, shell-crossing can occur at less negative $\delta_{\rm E, sc}$ than in GR, and the relative shift with respect to EdS can act as a distinctive MG signature rather than a pure background-expansion effect. 

Finally, we also showed how to adapt our hydrodynamical implementation to match the \textit{ad hoc} regularization adopted in Galileon $N$-body simulations~\cite{Barreira:2013eea}, where models that would otherwise be discarded by the void-informed criterion are instead evolved by enforcing a real-valued branch. In this controlled test, we intentionally considered parameter choices that violate the void-informed bound and selected void trajectories that would enter the pathological regime in the absence of regularization. We then showed how the prescription modifies the corresponding void solutions, tracing the transition from the physical to the regularized regime in both $\mu_{\rm NL}$ and $\delta_{\rm E}$.

Overall, within the controlled setup considered here (spherical symmetry, inverse top-hat and QSA), our results show that cosmic voids provide a particularly sharp probe of late-time departures from GR in theories with derivative self-interactions: they are simultaneously sensitive to the enhanced effective coupling and to the consistency of the non-linear branch structure. The hydrodynamical framework developed here makes these connections explicit by linking the EFT functions to Eulerian void evolution and to shell-crossing, while the void-informed viability criterion offers a complementary, physically motivated filter on parameter space. Taken together, these ingredients establish a self-consistent baseline for interpreting void dynamics in a broad class of MG theories. 

\acknowledgments
T.M. and N.F. acknowledge Baojiu Li for valuable inputs and insightful comments.
T.M. thanks Philip Mocz and Jeff Jennings for valuable discussions that helped improve the numerical code. T.M. acknowledges financial support from the Italian Space Agency (ASI) through the ASI-CAIF fellowship.
T.M., N.F., and F.P. acknowledge  the COST Action CosmoVerse, CA21136, supported by COST (European Cooperation in Science and Technology). T.M. and N.F. acknowledge the Istituto Nazionale di Fisica Nucleare (INFN) Sez. di Napoli, Iniziativa Specifica InDark.
T.M. and G.V. acknowledge support from the Simons Foundation to the Center for Computational Astrophysics at the Flatiron Institute. T.M. acknowledges the COST Action COSMIC WISPers, CA21106, and the COST Action BridgeQG, CA23130.
F.P. acknowledges partial support from the INFN grant InDark and from the Italian Ministry of University and Research (\textsc{mur}), PRIN 2022 `EXSKALIBUR – Euclid-Cross-SKA: Likelihood Inference Building for universe's Research', Grant No.\ 20222BBYB9, CUP C53D2300131 0006, and from the European Union -- Next Generation EU.

\appendix

\section{Three diagnostics}\label{sec:three_diagnostics}
In this section, we describe how we implement the computation of the three quantities introduced in section~\ref{sec:viability}.

We begin with $\delta_{\rm min,p}$. For each fixed target redshift $\bar{z}$, we scan over the ICs $\delta_{\rm v,in}$ (see eq.~\eqref{eq:ICs_hydrodynamical_eqs}) and identify the most negative value that still yields a real evolution up to $\bar{z}$, meaning that the solution remains real at all redshifts $z\le\bar{z}$. We implement this via a shooting strategy over a fixed IC interval, starting from two well separated values, $\delta_{\rm v,in}=-10^{-10}$ and $\delta_{\rm v,in}=-10^{-2}$. These bounds are chosen to bracket a sufficiently wide range for root finding, and we have explicitly verified that widening the interval does not affect the inferred value of $\delta_{\rm min,p}$. We then define an indicator function that tests the reality of the evolution up to $\bar{z}$,
\begin{align}
    s(\delta_{\rm v,in},\bar{z})=
    \begin{cases}
    +1\,, & \text{real up to } \bar{z}\,,\\
    -1\,, & \text{imaginary before } \bar{z}\,,
    \end{cases}
\end{align}
and use bisection on $\delta_{\rm v,in}$ until the critical initial condition is determined to a relative precision of $10^{-3}\%$. If the deepest tested IC still produces a real evolution, then no pathological onset is found within the explored IC range. Given the extremely deep initial underdensity adopted, the evolution is driven arbitrarily close to the physical bound $\delta_{\rm E}\to -1$, and we therefore set $\delta_{\rm min,p}(\bar{z})=-1$. Since $\delta_{\rm min,p}$ is defined without enforcing shell-crossing, it can correspond to depths that would be inaccessible once shell-crossing is imposed, which is consistent with its purely diagnostic role.

We now turn to $\delta_{\rm min,h}$. Its computation proceeds in two steps and requires $\delta_{\rm min,p}$ as an input. First, we determine the shell-crossing threshold by evolving the void while enforcing the prescription that prevents the square root in eq.~\eqref{eq:prescription_N_body_imaginary_force} from becoming imaginary. This prescription provides a well defined continuation for $\mu_{\rm NL}$ and does not halt the dynamics. The void solution keeps expanding and eventually undergoes shell-crossing. Second, we set $\delta_{\rm min,h}$ to the larger value between the shell-crossing bound and $\delta_{\rm min,p}$. This construction is consistent because, if the shell-crossing threshold is larger than $\delta_{\rm min,p}$, then shell-crossing occurs before the evolution enters the pathological branch, and the threshold is physically meaningful. Conversely, if the shell-crossing threshold is more negative than $\delta_{\rm min,p}$, then the solution becomes pathological first, so the evolution cannot be followed reliably down to shell-crossing in the original system, and the relevant bound is set by $\delta_{\rm min,p}$.

We finally turn to the third quantity, $\delta_{\rm min,p_0}(z)$. As already described in the main text, we first compute $\delta_{\rm min,p}(z=0)$ and determine the corresponding critical IC, $\delta_{\rm v,in,min,p_0}$, defined by
\begin{align}
        \delta_{\rm E}(z_{\rm in})=\delta_{\rm v,in,min,p_0}
        \quad\Rightarrow\quad
        \delta_{\rm E}(z=0)=\delta_{\rm min,p}(z=0)\,.
\end{align}
We then evolve the hydrodynamical equations starting from $\delta_{\rm v,in,min,p_0}$ and record the resulting trajectory as a function of $z$. The value of this trajectory at each redshift defines $\delta_{\rm min,p_0}(z)$.

\section{Non-linear void dynamics with linear gravitational coupling}\label{Sec:non_linear_void_dynamics_with_linear_gravitational_coupling}

In this appendix, we address a simple but instructive question: how does the evolution of $\delta_{\rm E}$ change if, in the non-linear evolution equation eq.~\eqref{eq:delta_nl_MG_rewrite}, we replace the non-linear gravitational coupling $\mu_{\rm NL}$ with its linear counterpart $\mu_{\rm L}$, while keeping the same background cosmology. We quantify the impact by computing the relative percentage difference between the two solutions, defined as
\begin{align}
    \Delta\delta_{\rm E}[\%] \,\equiv\,
    \frac{\delta_{\rm E}^{\mu_{\rm L}}-\delta_{\rm E}^{\mu_{\rm NL}}}{\delta_{\rm E}^{\mu_{\rm NL}}}\times 100\,,
\end{align}
where $\delta_{\rm E}^{\mu_{\rm NL}}$ denotes the non-linear density evolution obtained with the full coupling $\mu_{\rm NL}$, while $\delta_{\rm E}^{\mu_{\rm L}}$ is the solution obtained by substituting $\mu_{\rm NL}\to\mu_{\rm L}$ in eq.~\eqref{eq:delta_nl_MG_rewrite}. In all cases, we adopt the same ICs, calibrated so that the reference $w_0w_a$CDM evolution satisfies $\delta_{\rm E}(z=0)=-0.5$.

\begin{figure}
    \centering
    \includegraphics[width=1.0\linewidth]{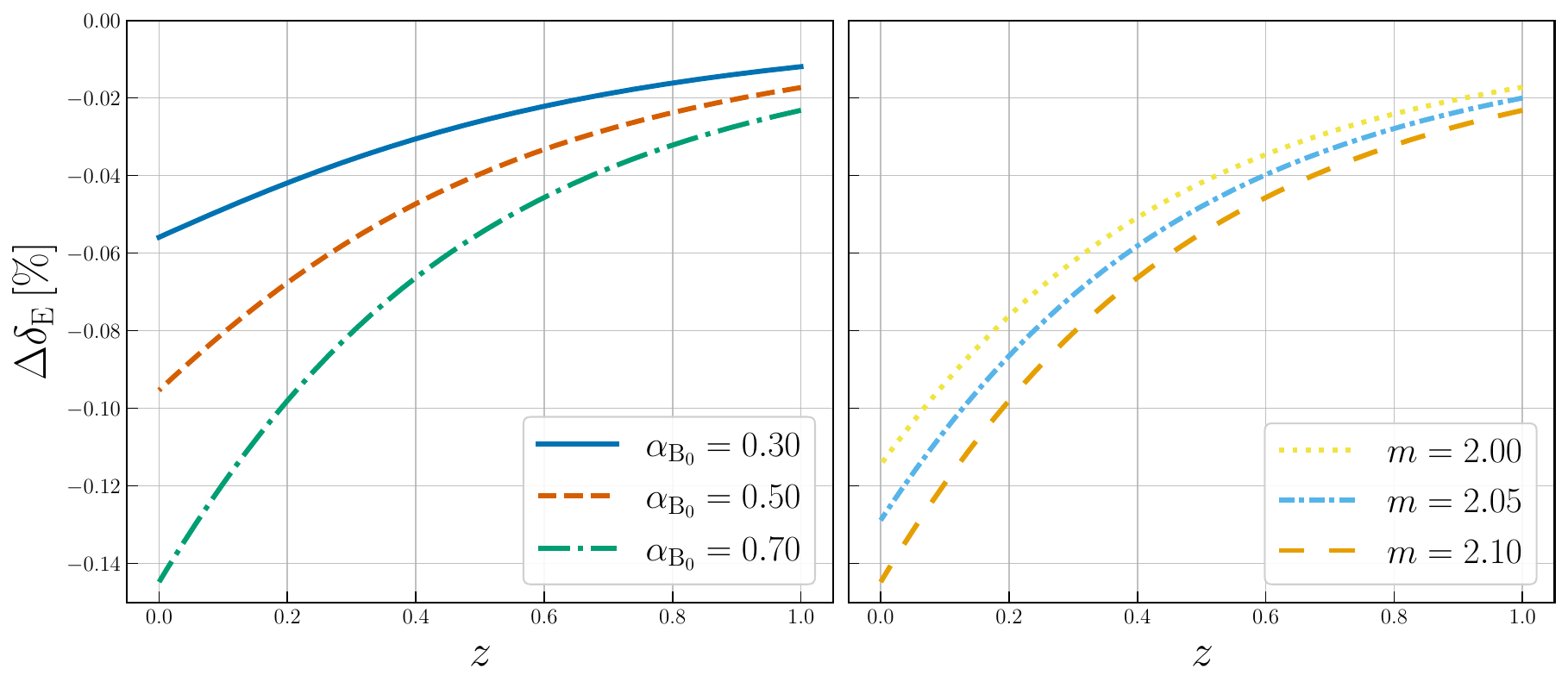}
    \caption{Relative percentage difference $\Delta\delta_{\rm E}[\%]$ between the density evolution obtained by replacing $\mu_{\rm NL}$ with $\mu_{\rm L}$ in eq.~\eqref{eq:delta_nl_MG_rewrite} and the solution computed with $\mu_{\rm NL}$, shown over $z\in[0,1]$ for the background parameters in eq.~\eqref{eq:baseline_background}. Left panel: varying $\alpha_{\rm B_0}\in\{0.3,0.5,0.7\}$ at fixed $m=2.1$. Right panel: varying $m\in\{2.00,2.05,2.10\}$ at fixed $\alpha_{\rm B_0}=0.7$.}
    \label{fig:single_solution_percent_diff_mu_lin_vs_mu_non_lin}
\end{figure}
The result is shown in figure~\ref{fig:single_solution_percent_diff_mu_lin_vs_mu_non_lin}, over $z\in[0,1]$ and for the baseline background parameters in eq.~\eqref{eq:baseline_background}.  The right panel varies $\alpha_{\rm B_0}\in\{0.3,0.5,0.7\}$ at fixed $m=2.1$, while the left panel varies $m\in\{2.00,2.05,2.10\}$ at fixed $\alpha_{\rm B_0}=0.7$.

Across the full range shown, the deviation remains subpercent. Nevertheless, it is systematic and shows that replacing $\mu_{\rm NL}$ with $\mu_{\rm L}$ is not a consistent approximation once subpercent accuracy is targeted. The difference is always negative, implying $\delta_{\rm E}^{\mu_{\rm L}}>\delta_{\rm E}^{\mu_{\rm NL}}$. This sign is expected because substituting $\mu_{\rm NL}\to\mu_{\rm L}$ weakens the source term in eq.~\eqref{eq:delta_nl_MG_rewrite} and, therefore, underestimates the efficiency of void evacuation. The magnitude of the effect increases toward low redshift, where voids become more unscreened and the discrepancy between $\mu_{\rm NL}$ and $\mu_{\rm L}$ grows as the void deepens. Finally, the parameter dependence follows the same trend discussed throughout: increasing either $\alpha_{\rm B_0}$ or $m$ enlarges the separation between $\mu_{\rm L}$ and $\mu_{\rm NL}$ and, consequently, amplifies the difference between the two evolutions.

\newpage

\bibliographystyle{JHEP}
\bibliography{main.bib}

\end{document}